\documentclass[%
 prb
 amsmath,amssymb,
 reprint,%
]{revtex4-2}

\usepackage{amsmath}
\usepackage{amssymb}
\usepackage{graphicx}
\usepackage{caption}
\usepackage{subcaption}

\usepackage{color}
\usepackage{tabularx}
\usepackage{dcolumn} % Align table columns on decimal point
\usepackage{bm}      % bold math
\usepackage{array}   % table making
\usepackage{amsfonts}
\usepackage{url}                     % Moved up here
\usepackage[bookmarks=false,colorlinks,citecolor=red]{hyperref} % Moved up here

\usepackage{placeins}
\usepackage{ulem}

\usepackage[utf8]{inputenc}
\usepackage[T1]{fontenc}
\usepackage{mathptmx}

\include{MyCommand}

\usepackage[ansiapaper]{geometry}
\usepackage{graphicx}
\usepackage{amsmath}
\usepackage{comment}
\usepackage{amsmath,cases}
\usepackage{gensymb}
\usepackage{afterpage}
\usepackage{braket}
\usepackage{physics}
\usepackage[toc,page]{appendix}
\usepackage{float}
\usepackage[inkscapelatex=false]{svg}


\begin{document}

\author{Kenneth O. Berard}
\affiliation{Department of Chemistry, Brown University, Providence, Rhode Island 02912, USA}

\author{Brenda Rubenstein}
\affiliation{Department of Chemistry, Brown University, Providence, Rhode Island 02912, USA}
\affiliation{Department of Physics, Brown University, Providence, Rhode Island 02912, USA}
\affiliation{Data Science Institute, Brown University, Providence, Rhode Island 02912, USA}

\author{Jaron T. Krogel}
\email{krogeljt@ornl.gov}
\affiliation{Materials Science and Technology Division, Oak Ridge National Laboratory, Oak Ridge, TN 37831, United States}

\begin{titlepage}
%\title{Denoising Electron Densities Generated by Quantum Monte Carlo Methods for Machine Learning Applications}
%\title{Information-Theoretic Analysis of DMC Density Denoising: From Fourier Filters to 3D UNETs}
\title{Denoising Diffusion Monte Carlo Electron Densities with Physically Informed Variance Stabilization: From Fourier Filters to 3D UNETs}
%\title{Title}

\thanks{
  Notice: This manuscript has been authored by UT-Battelle, LLC, under contract DE-AC05-00OR22725 with the US Department of Energy (DOE). The US government retains and the publisher, by accepting the article for publication, acknowledges that the US government retains a nonexclusive, paid-up, irrevocable, worldwide license to publish or reproduce the published form of this manuscript, or allow others to do so, for US government purposes. DOE will provide public access to these results of federally sponsored research in accordance with the DOE Public Access Plan (\url{http://energy.gov/downloads/doe-public-access-plan}).
}

%BR propositions: \title{Denoising Quantum Monte Carlo Electron Densities...for Accelerating Statistical Convergence and Machine Learning Applications} 
\begin{abstract}
\begin{center}
    \includegraphics[width=0.5\textwidth]{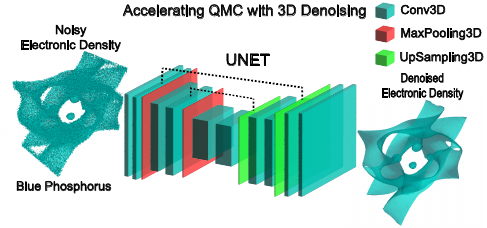}
    % Optional: Add a caption using \captionof{figure}{Your Caption}
    % \label{fig:Slice_comp}
\end{center}
\vspace{1em} % Adds a small space between the image and the abstract text
    \par
    Obtaining accurate electron densities is important for the fundamental description of molecular and condensed matter systems, as well as for the development of next-generation density functionals. Diffusion Monte Carlo (DMC), in particular, is known to produce benchmark-quality data; however, the predicted real-space electron densities contain substantial amounts of statistical noise. In this work, we study denoising approaches for DMC densities, judged on the basis of the information-theoretic Jensen-Shannon divergence. The denoising is facilitated by an approximate heteroscedastic to homoscedastic transformation leveraging the density functional theory density as a physical prior.     
    We systematically compare a range of denoising techniques---including Fourier transform, regression, and 3D UNETs---on materials showing a wide range of density variations: carbon diamond, blue phosphorus, and rutile VO$_2$. Our results indicate that simple flattened machine learning models and 2D image-based models introduce line artifacts and struggle to capture the full spatial correlation. In contrast, when using variance stabilization, regression methods outperform all others in both the high and low-noise limits across all materials considered. The best denoisers reduce the required cost of density-generating DMC simulations by 10--100$\times$, providing a promising route forward for application in noise-sensitive tasks such as DFT functional inversion. 
    %Finally, we develop automatic hyperparameter selection for a preferred regression method based on both pre-tuning and \textit{in situ} search for practical use in denoising stochastic electron densities for downstream machine learning and functional inversion applications.

\end{abstract}
\maketitle
\end{titlepage}

\setcounter{secnumdepth}{1}
\clearpage
\section{Introduction}
Electron densities play a pivotal role in our understanding of molecules and quantum materials.\cite{HohenbergKohn1964} By indicating the likelihood that electrons may be found at different positions in space, electron densities provide valuable insights into the types and strengths of bonds, chemical reactivity,\cite{Bader1990}  and the phase behavior of complex solids such as magnets and correlated metals.\cite{doi:10.1126/science.1107559,RevModPhys.70.1039}

Experimentally, the density of electrons around nuclei can be determined using advanced X-ray diffraction, electron scattering, and other spectroscopic techniques.\cite{Coppens1998,Genoni2018} While invaluable for extracting the bulk thermodynamic and electronic properties of materials, high-accuracy experimental determinations of the density are fraught with challenges. They are exceptionally expensive, highly dependent on competitive beamtime access at global synchrotron facilities, and often struggle with the well-known crystallographic phase problem or the resolution of transient states. Conversely, benchtop X-ray platforms offer broader accessibility, but suffer from inherently limited spatial resolution and accuracy, rendering them insufficient for mapping subtle correlation effects. 

Theoretically, electron densities are also key outputs of many electronic structure theories that enable direct comparisons with experiment and can thus provide pivotal insights into the performance of different theoretical methods. Density functional theory (DFT)\cite{HohenbergKohn1964,KohnSham1965} has historically served as the primary workhorse for calculating the spatial distribution of electrons. While DFT scales highly favorably with system size compared to wavefunction-based approaches for solving the many-body Schr\"{o}dinger equation, its absolute accuracy is fundamentally bottlenecked by the necessary approximation of the exchange-correlation (XC) functional. As modern functional approximations have ascended Jacob's ladder,\cite{Perdew2001} they have become increasingly complex and heavily parameterized to reproduce experimental thermodynamic and kinetic data. However, heavily empirical forms often stray from enforcing the underlying exact quantum mechanical constraints they were initially designed to capture, yielding artificially distorted electron densities even when energetic predictions appear accurate.\cite{Kepp2017,Kim2013} Consequently, obtaining true, high-accuracy electron densities necessitates algorithms that explicitly treat many-body electronic correlation without empirical bias.

Real-space quantum Monte Carlo methods, such as the diffusion Monte Carlo (DMC) method, have a long history of producing high-accuracy, many-body electron densities.\cite{Ceperley1980} The explicitly correlated physics introduced by Jastrow factors---which exactly enforce the electron-electron and electron-nucleus cusp conditions---and the imaginary-time projection of the ground state in DMC provide highly valuable, mathematically rigorous ground-truth density references.\cite{Foulkes2001,Kato1957} As a result, VMC and DMC have become useful tools for simulating the electronic properties of strongly-correlated solids whose sizes are often beyond the reach of other high-accuracy methods, such as coupled-cluster theory.\cite{10.1063/1.1727484} Nonetheless, while random sampling improves QMC's scaling, it results in statistical errors (noise) that accompany all QMC observables, electron densities included. Prior algorithmic efforts, such as zero-variance zero-bias (ZVZB) estimators, have been developed to mitigate this noise for various observables.\cite{Assaraf1999,Toulouse2007} However, because projecting data onto a dense 3D real-space grid inherently reintroduces spatial statistical fluctuations, isolating perfectly smooth electronic densities still ultimately requires exhaustive sampling.

This state of affairs often precludes the straightforward use of QMC densities for comparison to those produced by other computational methods or experiment and in different machine learning tasks. Because DMC falls among the relatively few real-space many-body methods with low-order polynomial scaling, the lack of smooth densities makes it difficult to leverage DMC's real-space advantage to assess basis set errors in other, orbital-based methods. Crucially, while data-driven machine learning models hold immense promise for predicting electronic properties, training these models requires massive datasets of highly accurate, continuous densities. Generating these datasets via brute-force QMC sampling is computationally intractable. Therefore, establishing a robust denoising surrogate is a strong prerequisite for leveraging QMC data in downstream machine learning tasks.

One particularly interesting modern use of electron densities is in developing more exact exchange-correlation (XC) functionals for use in DFT. Two of the key theorems underpinning DFT are the Hohenberg-Kohn Theorems, which effectively state that exchange correlation functionals can be mapped one-to-one to ground state electron densities. These theories are popularly used in the forward direction in which exchange-correlation functionals are constructed to yield accurate densities. But, as has been recognized for nearly 50 years, the one-to-one mapping implies that electron densities can be inverted to learn accurate functionals. Recent works have attempted to use highly accurate electron densities produced using explicitly correlated methods to learn a more exact (if not exact) exchange-correlation functional. For example, numerical density-to-potential inversion techniques have been successfully applied to benchmark densities generated via full configuration interaction (FCI) and density matrix renormalization group (DMRG) theory to reverse-engineer exact correlation potentials \cite{Shi2021, Kanungo2019}. However, the steep polynomial or exponential scaling of these methods restricts the range of physical systems these methods can address, heavily bottlenecking the learning of high-quality functionals for complex extended solids.

Given this backdrop, a way to reduce the statistical noise on QMC electron densities would open up several avenues of research by providing smooth, highly accurate, and relatively large-scale material densities. In this work, we study denoising approaches as a means of reducing the statistical errors on QMC densities. Our underlying hypothesis is that advanced machine learning and numerical denoising algorithms can overcome the fundamental QMC sampling problem, rapidly producing high-quality continuous electron densities capable of guiding \textit{ab initio} functional development. 

%Just as experimental X-ray scattering data is mathematically inverted via Fourier transforms to yield real-space electron densities, related numerical inversions can be performed on highly accurate theoretical spatial densities to construct exact DFT exchange-correlation potentials.\cite{Chayes1985, WuYang2003, ZhaoMorrisonParr1994, Kanungo2025, Jensen2018, Shi2021,ravindran2025densitiespotentialsbenchmarkinglocal} Rather than conventionally solving the Kohn-Sham equations to iteratively find the density from an approximated potential, these density-to-potential inversion techniques take a target real-space density and strictly reverse-engineer the unique corresponding effective Kohn-Sham potential. By isolating the exact XC potential from this inversion, researchers can directly map the exact functional derivative, bypassing decades of empirical guesswork. Naturally, this reverse-mapping relies heavily on the gradients and Laplacian of the density. It is therefore mathematically ill-posed and practically unstable if the input QMC density is corrupted by excessive stochastic noise; even minor statistical fluctuations result in wildly unphysical, divergent potentials,\cite{Bulat2007,Kanungo2019,Ospadov2017} underscoring the absolute necessity for highly refined, rigorously smooth input data.

Fortunately, there is a rich algorithmic history dedicated to mitigating statistical noise in spatial data. Within the physical and medical sciences, a variety of sophisticated mathematical formulations, such as total variation (TV) denoising \cite{Rudin1992}, Block-Matching and 3D filtering (BM3D) \cite{Dabov2007}, and Block-Matching and 4D filtering (BM4D) \cite{maggioni2013} have been engineered to process noisy signals generated by experimental instrumentation like magnetic resonance imaging (MRI) and computed tomography (CT) scanners. Most historical work in this domain has heavily focused on standard 2D image processing. In recent years, deep learning approaches, particularly convolutional neural networks (CNNs) and related UNET architectures, have demonstrated state-of-the-art, super-resolution performance on 2D image data.\cite{Ronneberger2015,Cicek2016} The UNET's encoder-decoder structure, fortified with skip connections, is theoretically ideal for preserving high-frequency spatial features while successfully smoothing low-frequency background noise. However, the direct translation of these 2D vision models to our 3D quantum chemical densities is severely hindered by the fundamental physical nature of spatial correlation. In 3D electron densities, the physical correlation between adjacent voxels is governed by complex orbital decay, angular momentum, and multi-reference interactions, rather than macroscopic visual features.\cite{Schutt2017} Furthermore, electronic densities span several orders of magnitude across a single simulation cell, making the data structure vastly more subtle and mathematically rigid than what can be treated accurately by simply coercing it into a standard 2D image format.

Building upon this foundational work in signal processing and deep learning, we have designed and evaluated a suite of denoising techniques tailored specifically for the rigorous demands of 3D quantum electron densities. To comprehensively benchmark these methods, we selected three chemically distinct systems: carbon diamond, blue phosphorus, and vanadium dioxide. These systems represent varying degrees of physical complexity and algorithmic difficulty. Carbon diamond provides a baseline ``easy'' case characterized by smoothly varying, well-defined covalent bonds and high crystallographic symmetry. Blue phosphorus represents a ``medium'' challenge: while its 2D in-plane electronic density is relatively straightforward to resolve, capturing its long, exponentially decaying evanescent vacuum tails without accidentally truncating or heavily distorting the distribution is highly difficult \cite{Tal1978}. Finally, vanadium dioxide presents a ``hard'' case governed by strong electronic correlation and Mott insulator physics,\cite{Biermann2007,Zheng2015} distinguished by complex spatial features arising from highly localized, high-density transition metal d-orbitals interacting with diffuse oxygen p-orbitals. 

\begin{figure*}[htbp]
    \centering
    \includegraphics[width=0.9\linewidth]{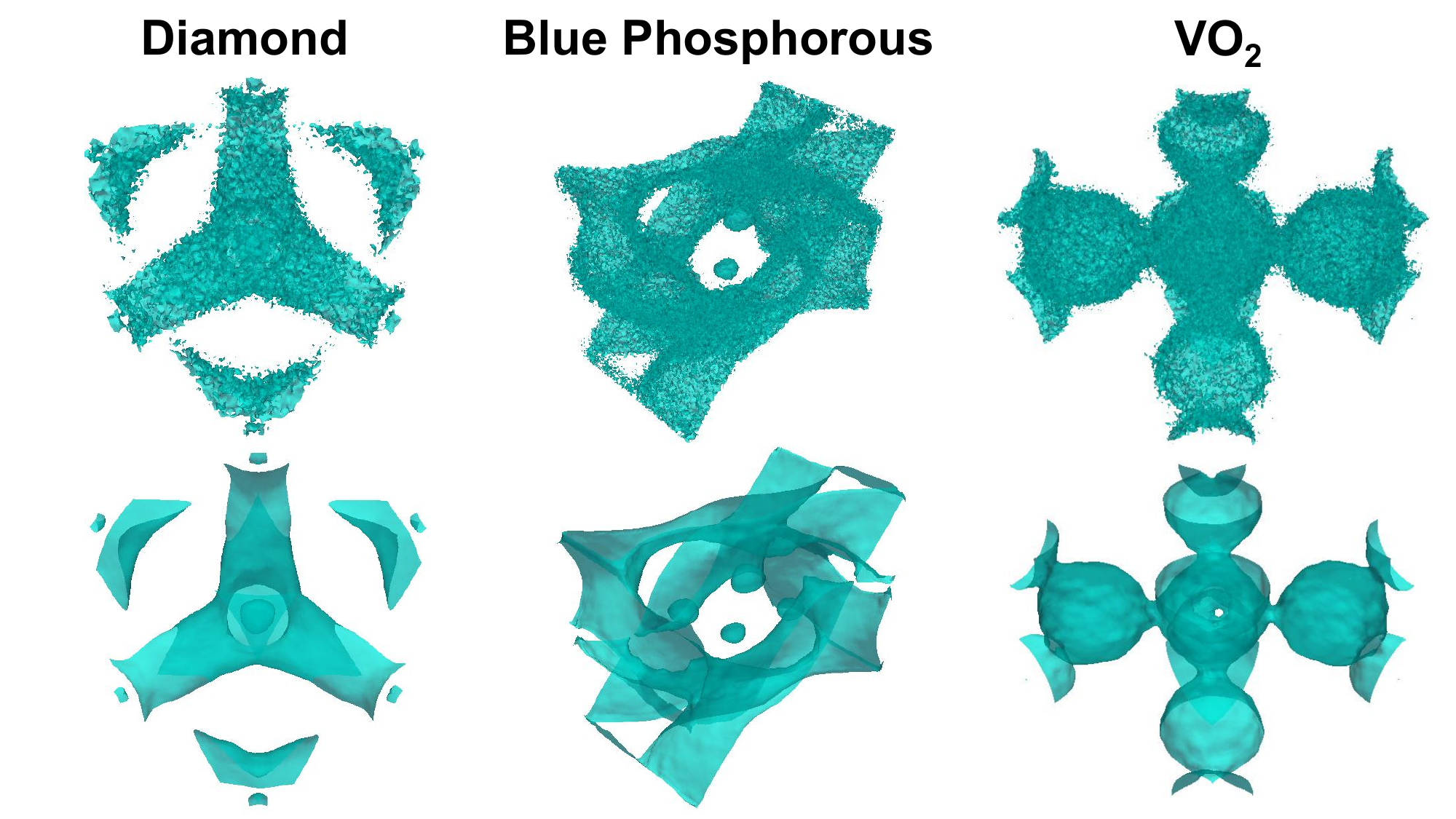}
    \caption{Denoised isosurfaces of low-sample DMC electron densities for diamond, blue phosphorus, and VO$_2$. The top row displays the raw, highly noisy DMC densities, while the bottom row shows the smooth, physically coherent isosurfaces recovered using the 3D UNET.}
    \label{fig:Iso_vis}
\end{figure*}

To visually demonstrate the efficacy of this approach on real electron density data, Figure~\ref{fig:Iso_vis} illustrates the 3D electron density isosurfaces for these three materials before and after applying our 3D UNET denoising procedure. The stochastic noise---which heavily obscures the true spatial distribution in low-sample QMC---is visibly eliminated without hallucinating artificial density or destroying valid physical signals. The denoised isosurfaces clearly resolve fine physical details that are otherwise indistinguishable in the noisy versions; for example, a central sphere becomes clearly resolved in carbon diamond and distinct topological features become apparent in both blue phosphorus and VO$_2$.

In this work, noisy 3D densities for these systems were generated using both VMC and DMC with varied Jastrow factors. These noisy densities were subsequently processed using a spectrum of machine learning and numerical denoising methodologies. Overall, algorithmic performance was systematically gauged by the achieved effective computational speedup related to the statistical and systematic fidelity of the reconstructed densities.  This was enabled by using the Jensen-Shannon Divergence (JSD),\cite{Lin1991} an information-theoretic measure of probability distribution overlap. 

To directly benchmark against off-the-shelf image processing models, we initially encoded the 3D densities into pseudo-images, mapping discrete 2D slices along the z-axis to independent color channels. We then rigorously compared these to custom denoisers that explicitly preserve and operate upon the native 3D physical structure. Our results demonstrate that massive computational speedups---effectively circumventing millions of required Monte Carlo steps---can be achieved using these models without the need to leverage massive swaths of costly QMC training data. While numerical 2D image denoising models exhibited moderately beneficial performance, they were outperformed by explicitly trained, mathematically tailored 3D architectures. Ultimately, our specialized 3D pipelines produce smooth, physically accurate electron densities that are statistically representative of QMC simulations requiring orders of magnitude more sampling than the underlying noisy input, paving a viable pathway for the routine extraction of QMC densities for exact functional development and other electronic structure applications.

\section{Theoretical Background and Data Pre-processing}

\subsection{QMC Electron Densities as Histograms}
The exact definition of the electron density is 
\begin{equation}
  \rho(\mathbf{r}) = \int d\mathbf{R}~P(\mathbf{R}) \sum_{i=1}^{N_e}\delta(\mathbf{r}-\mathbf{r}_i),
\end{equation}
where $N_e$ is the electron count, $\mathbf{r}_i$ is the coordinate of the $i$-th 
electron, $\mathbf{R}=[\mathbf{r}_1\cdots \mathbf{r}_{N_e}]$ is the joint coordinate of all of the electrons, 
and $P(\mathbf{R})$ is the sampling distribution ($P=|\Psi_T|^2$ for VMC and 
$P=|\Psi_T\Psi_0|$ for DMC).

The QMC electron density is commonly represented as a volumetric histogram:
\begin{equation}
    \rho(\mathbf{r}_m) = \int_{\Omega_m} d\mathbf{r} \rho(\mathbf{r}).
\end{equation}
Here, $\Omega_m$ represents voxel $m$ centered at $\mathbf{r}_m$, 
with the histogram grid containing $M$ total cells. 
The density is normalized to the electron count: $\sum_{m=1}^M\rho_m = N_e$.
In the discussion that follows, we use $\rho(\mathbf{r}_m)$ and $\rho_m$ interchangeably.

\subsection{Poisson Model for Noisy Electron Densities}
\label{sec:noise_model}

The Poisson distribution accurately reflects situations in which independent events occur at a constant localized rate. The form of the distribution is
\begin{equation}
    \mathcal{P}_{pois}(n;\mu) = \frac{\mu^n e^{-\mu}}{n!}
\end{equation}
where $\mu$ is the mean number of occurrences and $\sum_{n=0}^\infty\mathcal{P}_{pois}(n;\mu)=1$. For a stochastic variable $\eta$ sampled from $\mathcal{P}_{pois}(\mu)$ (denoted $\eta\sim\mathcal{P}_{pois}(\mu)$), the expected value is $\langle\eta\rangle=\mu$.

In a density histogram, the mean number of occurrences in each cell is proportional to the density within that voxel. The entire density histogram can be formulated as a probability distribution via renormalization, $p(\mathbf{r}_m)=\rho(\mathbf{r}_m)/N_e$. If we consider $N$ total samples drawn from this 3D distribution, the average count falling in a particular histogram cell is $Np_m$. If a QMC random walk has accumulated $N_{mc}$ walker configurations, the total number of single-particle samples is $N=N_{mc}N_e$, and the expected count is $Np_m=N_{mc}N_ep_m=N_{mc}\rho_m$. Here, $\rho_m$ represents the true, noise-free electron density in that cell.

The histogram counts ($c$) are distributed in each respective cell as:
\begin{equation}
    \mathcal{P}_{count}(c;N_{mc},m) = \mathcal{P}_{pois}(c;N_{mc}\rho_m)
\end{equation}
Therefore, if a random variate $\tilde{c}_m$ is drawn from the electron count distribution in cell $m$, i.e., $\tilde{c}_m\sim\mathcal{P}_{count}(N_{mc},m)$, then $\tilde{\rho}_m=\tilde{c}_m/N_{mc}$ follows the corresponding electron density distribution, $\tilde{\rho}_m\sim\mathcal{P}_{dens}(\rho;N_{mc},m)$ and has a mean of $\langle\tilde{\rho}_m\rangle=\langle\tilde{c}_m\rangle/N_{mc}=N_{mc}\rho_m/N_{mc}=\rho_m$. 
When random samples are drawn from $\mathcal{P}_{dens}$ for each cell, the resulting 3D density represents a ''single sample'' of the density histograms possible from Monte Carlo runs gathering $N_{mc}$ configurations.

In the high-sampling/low-noise limit $(N_{mc}\xrightarrow{}\infty)$, Poisson distributions approximate a Gaussian (normal) distribution. Denoting a normal distribution with mean $\mu$ and standard deviation $\sigma$ as $\mathcal{N}(\mu,\sigma^2)$, a Poisson distribution with a large mean asymptotically approaches $\lim_{\mu\to\infty}\mathcal{P}_{pois}(\mu)=\mathcal{N}(\mu,\mu)$. Therefore, in the high-sampling limit, the random variates for electron count and density in histogram cell $m$ are defined as:
\begin{align}
    \tilde{c}_m &= N_s\rho_m + \sqrt{N_{mc}\rho_m}\eta_m \\
    \tilde{\rho}_m &= \rho_m + \sqrt{\frac{\rho_m}{N_{mc}}}\eta_m
\end{align}
where each $\eta_m$ is drawn from a unit normal distribution.

\subsection{Invertible Density Transformations to Facilitate Denoising}\label{sec:transforms}

Variable scales in both the signal and noise in noisy electron densities pose significant challenges for denoising techniques. Electron densities can span multiple orders of magnitude within a given material while also  potentially exhibiting areas of rapidly varying localized charge (such as in 3d shells) and slowly varying dispersed charge (such as in evanescent regions extending into van der Waals gaps). As seen from the models in section \ref{sec:image_results}, the noise also varies drastically in different regions, with low density regions in particular exhibiting small signal-to-noise ratios. Ultimately, the performance of a denoiser will depend on its ability to reduce noise across the material without introducing much deformation in the density itself. From this point of view, invertible transformations of the density can help to reduce the variability in the density signal, while also potentially evening out the scale of the noise.

\paragraph{Square Root Transformation:}
Because QMC statistical fluctuations follow a Poisson-like distribution---where the local variance scales proportionally with the expected electron density---taking the square root serves as a classical variance-stabilizing transformation. This mathematically compresses the extreme dynamic range of the core electron density. To ensure strict numerical stability, the absolute value is taken prior to transformation. The transformed input $u(\mathbf{r})$ is
%and the corresponding true target $y(\mathbf{r})$ are defined as:
\begin{align}
u_\mathrm{noisy}(\mathbf{r}) &= \sqrt{|\rho_{\text{noisy}}(\mathbf{r})|} 
%y(\mathbf{r}) &= \sqrt{|\rho_{\text{true}}(\mathbf{r})|}
\end{align}
During inference, the physically continuous target density is exactly reconstructed by squaring the network's predicted output: $\rho_\mathrm{clean}(\mathbf{r}) = \left(u_\mathrm{clean}(\mathbf{r})\right)^2$.

\paragraph{Approximate Homoscedastic Transformation Based on Physical Residuals:}
This transformation makes use of the fact that the DFT density resembles the QMC density in many respects, as each carries physical information about the system.
Therefore, the residual $\rho_\mathrm{noisy}(\mathbf{r})-\rho_\mathrm{DFT}(\mathbf{r})$ will be more slowly varying than the noisy density alone. 
In addition to this, the spatial variations in the noise (heteroscedasticity) can be mapped to nearly uniform Gaussian distributions with unit width (homoscedasticity) by dividing by the square root of the DFT density similar to a Z-score.
The transformed input $u_\mathrm{noisy}(\mathbf{r})$ is
\begin{equation}
    \quad u_\mathrm{noisy} (\mathbf{r})= \frac{\rho_\mathrm{noisy}(\mathbf{r})-\rho_{DFT}(\mathbf{r})}{ \sqrt{\rho_{DFT}(\mathbf{r})}+\epsilon}
\end{equation}
Denoising in this space yields $u_\mathrm{clean}(\mathbf{r})$.  Following this, the denoised density is found via the inverse transformation:
\begin{equation}
\rho_\mathrm{clean}(\mathbf{r}) = \rho_{DFT}(\mathbf{r}) + \left(\sqrt{\rho_{DFT}(\mathbf{r})}+\epsilon\right) u_\mathrm{clean}(\mathbf{r})
\end{equation}
Here, $\epsilon$ is a small numerical stabilizer that becomes unnecessary once a modest sampling level has been reached.

In the limit that the DFT density becomes exact, the task---from the point of view of regression---is to fit a flat hyperplane at zero to a uniform gaussian noise field.
Since most denoisers and regressors developed in the literature assume uniform Gaussian random noise, this transformation maps the density denoising problem into a space where these techniques are naturally most effective. 
%The effect of each of these transformations is shown in figure \xtask{}, using the VO$_2$ DMC density as an example. 
As we will show later, this transformation is key to identify and remove noise---without introducing substantial systematic bias---in the low-noise limit that is characteristic of real DMC calculations.

\subsection{Error Metrics and Quality Evaluation}
\label{sec:Error_Qual}
To quantitatively assess the fidelity of the denoised electron densities relative to the highly converged reference data, we require a rigorous statistical divergence metric. While the Kullback-Leibler (KL) divergence is a standard measure of relative entropy, it is inherently asymmetric and unbounded, making it difficult to establish an absolute scale of spatial convergence. Furthermore, standard KL divergence is physically ill-suited for electronic densities; because the exact physical density decays exponentially to near-zero in vacuum regions, the logarithmic ratio in the KL formulation asymptotically diverges, leading to severe numerical instability. 

We instead employ the Jensen-Shannon Divergence (JSD), a symmetrized and smoothed modification of the KL divergence. For discrete probability distributions $p_1$ and $p_2$ evaluated on a spatial grid, the JSD is defined as:
\begin{equation}\label{eq:djs}
  D_{JS}(p_1,p_2) = \frac{1}{2\log{2}}\sum_{m=1}^M\bigg[p_1\log{\frac{p_1}{p_{\text{mix}}}} + 
p_2\log{\frac{p_2}{p_{\text{mix}}}}\bigg]
\end{equation}
where $p_{\text{mix}} \equiv (p_1+p_2)/2$ is the mixture distribution. Unlike the KL divergence, the mixture distribution ensures that the logarithmic terms remain finite. Consequently, the JSD is symmetric, strictly bounded between 0 and 1, and serves as a reliable metric to evaluate the spatial proximity of the denoised density to the exact target distribution.

The high sampling asymptotics of the Jensen-Shannon Divergence play an important role in estimating the proximity of a denoised density to the full noise-free limit.  This is necessary to estimate the actual effective speedup gained by the denoising methods.  As we derive in supplementary section \ref{sec:asymp_deriv}, the JSD between two noisy distributions with the same underlying noise-free limit is
\begin{equation}\label{eq:div_asymp_pair}
    D_{JS}(p_1,p_2) \approx \frac{1}{8\log{2}}\left(\frac{M}{N_1}+\frac{M}{N_2}\right)
\end{equation}
where $M$ is the number of histogram cells and $N_1$ and $N_2$ are the number of samples used to create the respective noisy distributions.  The asymptotic divergence between a given distribution and its noise-free limit is then 
\begin{equation}\label{eq:div_asymp_single}
    D_{JS}(p_\mathrm{noisy},p_\mathrm{exact}) \approx \frac{1}{8\log{2}}\frac{M}{N}
\end{equation}
Importantly, the asymptotics are entirely independent of the distribution in question. 
This allows us to correct the divergence computed between a denoised density $\rho_\mathrm{DN}$ and a low-noise reference, where $N_{mc}$ Monte Carlo samples were used to create the noisy input density:
\begin{equation}\label{eq:div_corr}
    D_{JS}(p_\mathrm{DN},p_\mathrm{exact})\approx  D_{JS} (p_\mathrm{DN},p_\mathrm{noisy}) - \frac{1}{8\log{2}}\frac{M}{N_{mc}}
\end{equation}
These forms become good approximations at moderate sample counts, enabling their practical use in typical DMC calculations.

\subsection{Training Data Generation: A Synthetic Density Pipeline}

To efficiently generate robust training datasets and optimize model hyperparameters without relying on expensive Markov chain sampling or external DFT baseline calculations, we developed a fully synthetic density generation pipeline. This pipeline generates idealized target densities and subjects them to simulated QMC-like heteroscedastic noise. The process consists of three steps: synthetic initialization, physics-aware perturbation, and Poisson noise injection.

We first simulate an idealized electron distribution by placing three 3D Gaussian functions, representing pseudo-atomic centers, within a discrete volumetric grid. 
%Let the grid dimensions be $D_z, D_y, D_x$, defined by the coordinate vector $\mathbf{r} = (z, y, x)$. The locations of the three pseudo-atomic centers, $\mathbf{c}_1, \mathbf{c}_2$, and $\mathbf{c}_3$, are:
%\begin{align}
%\mathbf{c}_1 &= \left( \lfloor \frac{D_z}{2} \rfloor, \lfloor \frac{D_y}{2} \rfloor, \lfloor \frac{D_x}{2} \rfloor \right) \\
%\mathbf{c}_2 &= \left( \lfloor \frac{D_z}{3} \rfloor, \lfloor \frac{D_y}{2} \rfloor, \lfloor \frac{D_x}{2} %\rfloor \right) \\
%\mathbf{c}_3 &= \left( \lfloor \frac{D_z}{2} \rfloor, \lfloor \frac{D_y}{2} \rfloor + 15, \lfloor \frac{D_x}{2} %\rfloor \right)
%\end{align}
The raw synthetic reference density is generated by summing isotropic Gaussian functions (variance spread $\sigma = 6.0$~voxels) centered at these coordinates.
%\begin{equation}
%\rho_{\text{raw}}(\mathbf{r}) = \sum_{k=1}^{3} \exp\left( -\frac{\|\mathbf{r} - \mathbf{c}_k\|^2}{2\sigma^2} \right)
%\end{equation}
The specific Cartesian positions chosen for these pseudo-atomic centers are strictly arbitrary. This intentional geometric randomness ensures that the neural network learns the fundamental statistical distribution and heteroscedastic noise profile of the data, rather than introducing geometric bias or overfitting to specific molecular symmetries and bond lengths.

To simulate many-body electronic correlation and force the neural networks to learn diverse physical topologies rather than merely memorizing smooth Gaussian profiles, we introduce a physics-aware perturbation. We generate a smooth, low-frequency random continuous field $F(\mathbf{r}) \in [-1, 1]$ scaled by a random strength parameter $s \in [0.05, 0.15]$:
\begin{equation}
\tilde{\rho}(\mathbf{r}) = \rho_{\text{raw}}(\mathbf{r}) \left( 1 + s F(\mathbf{r}) \right).
\end{equation}
The resulting density is then purged of negative values and renormalized to the total electron count.

%To guarantee physical validity, two constraints are strictly enforced. First, a small non-negativity floor  $\delta$ 
%= 10^{-12}$ 
%is applied across the grid to prevent numerical instability during logarithmic or root variance transformations:
%\begin{equation}
%\tilde{\rho}_{\text{pos}}(\mathbf{r}) = \max(\tilde{\rho}(\mathbf{r}), \delta)
%\end{equation}
%Second, the grid is renormalized to ensure exact charge conservation, anchoring the total sum to a predefined target number of electrons, $N_e$:
%\begin{equation}
%\rho_{\text{target}}(\mathbf{r}) = \tilde{\rho}_{\text{pos}}(\mathbf{r}) \left( \frac{N_e}{\sum_{\mathbf{r}'} %\tilde{\rho}_{\text{pos}}(\mathbf{r}')} \right)
%\end{equation}
%This physically constrained, normalized spatial distribution, $\rho_{\text{target}}(\mathbf{r})$, serves as the exact ground-truth label for our supervised models.

%To comprehensively evaluate the robustness of the denoising architectures, the networks are trained and benchmarked across three distinct data representations: raw electronic densities, standardized residuals, and square-root transformed densities. 
%Training on the raw densities requires the network to natively navigate the extreme dynamic range and heteroscedastic noise profile of the physical signal. To make the optimization landscape more tractable, variance-stabilizing transformations can be applied to partially or fully decouple the local noise variance from the signal amplitude.

\subsection{Conversion of 3D Density to Images}

To leverage standard 2D image processing architectures, the 3D continuous volumes were encoded into pseudo-RGB image stacks. To manage the vast dynamic range while preserving the relative spatial structure of the electron density, the \textit{entire} 3D volume (dimensions $N \times 64 \times 64 \times 64$) was globally normalized to a $[0, 1]$ range using a single global min-max scaling factor, with a constant $\epsilon = 10^{-6}$ added to the denominator to prevent zero-division artifacts. 

The volume was decomposed along the $z$-axis into 64 discrete 2D slices. Each globally normalized 2D slice was subsequently duplicated across three identical color channels to synthesize a standard RGB format, yielding $N \times 64$ individual $64 \times 64 \times 3$ image tensors. Following the application of the 2D algorithms, the inverse mapping was performed: the first channel of each processed RGB image was extracted, and exact physical values were re-established by reverse-applying the global volume scaling factors.

\subsection{Materials and Datasets}
We chose three qualitatively different materials to investigate the 
properties of electron density denoising algorithms: carbon diamond ($a=3.57$~\AA{}), 
blue phosphorus (ICSD-25253), and R-phase VO$_2$ (ICSD-1504). 
Diamond represents a relatively easy case with a slowly varying density, 
blue phosphorus has low density vdW gaps between layers, and 
VO$_2$ contains a wide variation in density (high near the V $3d$ shell, 
intermediate near the oxygen $2p$, and low in interstitial regions).
In each case, we use a single primitive cell at the $\Gamma$-point. 
The electron density histogram grids are as follows: 
$64\times64\times64$ for diamond, 
$140\times84\times112$ for blue phosphorus, 
and $116\times116\times72$ for VO$_2$.
The width of the histogram cells is approximately 0.04~\AA{} in each 
dimension for all materials considered.

QMC electron density histograms were collected for all materials using both 
VMC and DMC with Slater-Jastrow trial wavefunctions.  VMC was also performed 
without a trial Jastrow-factor.  In that case, the noise-free limit of the 
electron density is identical to the DFT density.  Histograms were collected 
over the range of $3.6\times10^4~-~5.3\times10^9$ walker configurations in 
exponentially increasing intervals, including powers of two. 
DFT densities were interpolated via B-splines and integrated over $4\times4\times4$-centered grids in each QMC histogram cell to match the statistical binning process 
of the histograms.

\subsection{Computational Details}
All density functional theory calculations were performed with  Quantum ESPRESSO \cite{Giannozzi_2017,Giannozzi_2009}. Density functionals were chosen within the LDA family: pure LDA for closed shell and weakly-correlated diamond and blue phosphorus, and LDA+U for strongly correlated VO$_2$. A Hubbard U value of 3.5 eV was used in accordance with prior QMC  benchmark calculations of the VO$_2$ electron density \cite{Zheng2015}. Electronic densities were generated from extended QMCPACK runs using the Nexus workflow manager \cite{Kim_2018,QMCPACK_2,KROGEL2016154}. We note that, for the purpose of studying denoising, the accuracy of the density functional is not as important as in simulations meant to directly predict materials properties.

\section{Denoising Algorithms}

In this work, we employ a combination of classical transform-domain filters and modern deep learning architectures to isolate deterministic electronic density signals from stochastic QMC noise. 

\subsection{Block-Matching and Collaborative Filtering (BM3D and BM4D)}
To establish a rigorous classical baseline, we evaluate the Block-Matching and 3D filtering (BM3D) algorithm~\cite{Dabov2007} and its native volumetric extension, BM4D~\cite{maggioni2013}. Both approaches exploit non-local spatial self-similarity. BM3D groups structurally similar 2D cross-sectional density patches into 3D arrays, applying a 3D linear transform followed by hard-thresholding to truncate the uniformly distributed QMC noise. The transform is inverted to yield a local noise-attenuated estimate. These overlapping estimates are then aggregated into the original spatial coordinates via a weighted average, where the weights are inversely proportional to the residual noise variance, $w \propto (\sigma^2 N_{\text{nz}})^{-1}$ (with $N_{\text{nz}}$ representing the retained non-zero coefficients). This is refined in a second stage via empirical Wiener filtering. 

Because BM3D neglects out-of-plane physical correlations via 2D slicing, we predominantly rely on BM4D~\cite{maggioni2013}. Operating natively on 3D voxel cubes, BM4D stacks similar cubes into a 4D group. A separable 4D transform simultaneously exploits local 3D spatial correlations and non-local global similarities, natively preserving the isotropic nature of the density and avoiding slicing artifacts entirely.

\subsection{Fourier Transform Filtering and Augmentation}
Spectral filtering via the 3D Fast Fourier Transform (FFT) separates the low-frequency structural harmonics of the continuous density from broad-spectrum QMC shot noise. We project the noisy density into the frequency domain, $\tilde{\rho}(\mathbf{k})$, and dynamically construct a Boolean frequency mask by evaluating a variance-ceiling function, $f_{\text{ceil}}(j) = \max_{m \ge j} ( |\tilde{\rho}_{\text{noisy}}(m)| )$, on the spectral amplitudes. This precisely identifies the critical cutoff where the stochastic white-noise floor overtakes the exponentially decaying physical signal of the clean reference density. 

The noisy spectral components beyond this cutoff are either removed (zeroed) or replaced with the pristine reference amplitudes (augmented). Following the inverse 3D FFT, a strict positivity constraint, $\hat{\rho}(\mathbf{r}) = \max( 0, \mathcal{F}^{-1} \{ \tilde{\rho}_{\text{filtered}}(\mathbf{k}) \} )$, is enforced on the spatial grid to rectify localized Gibbs ringing.

\subsection{Regression: SmoothN Penalized Least Squares Smoothing Spline}
The SmoothN algorithm~\cite{garcia2010,garcia2011,WANG2012} provides a robust, non-parametric penalized least squares (PLS) smoothing framework. It computes a 3D smooth spline fit $\hat{y}(\mathbf{r})$ by minimizing a regularized energy functional:
\begin{equation}
    \mathcal{E}(\hat{y}) = \sum_{\mathbf{r}} w(\mathbf{r}) \left| y(\mathbf{r}) - \hat{y}(\mathbf{r}) \right|^2 + s \sum_{k=1}^{N} \left\| \Delta^2_k \hat{y}(\mathbf{r}) \right\|_2^2
\end{equation}
This balances data fidelity against a multi-dimensional second-order Laplacian penalty ($s$), heavily accelerated by the discrete cosine transform (DCT). To protect the field from heavy-tailed non-Gaussian noise, SmoothN employs an iteratively reweighted least squares (IRLS) M-estimation loop. The spatial weights $w(\mathbf{r})$ are dynamically updated using Tukey's bisquare objective function based on the Median Absolute Deviation (MAD) of the residual error field. This progressively assigns anomalous noise spikes a weight of zero, ensuring the output remains faithful to the true underlying physical distribution.

\subsection{Regression: Sequential Local 1D Polynomial Smoothing}
As a highly localized spatial baseline, we implement a moving-window 1D polynomial smoothing method. For each target voxel, an ordinary least squares polynomial $P(t) = \sum_{k=0}^{n} c_k t^k$ of degree $n$ is fitted to a localized 1D window of width $2w+1$. The smoothed scalar is evaluated precisely at the central coordinate, $P(w)$. This smoothing is applied sequentially along the $z$, $y$, and $x$ dimensions, utilizing periodic boundary wrapping to eliminate the need for external padding.

\subsection{Deep Learning: Convolutional Autoencoder and 3D Residual UNET}
Deep learning methods parameterize the mapping of noisy densities to clean target manifolds. While a standard Convolutional Autoencoder (CAE) effectively captures global dependencies, its strict spatial bottleneck irrevocably degrades the fine-grained physical details required for exact physical reconstructions. 

To natively process volumetric densities without this loss, we utilize a 3D Residual UNET. This architecture routes high-resolution spatial feature maps directly from the encoder to the symmetric decoder via skip connections, utilizing $3 \times 3 \times 3$ volumetric convolutions, max pooling, and nearest-neighbor upsampling. Crucially, the model maps the transformed noisy input to the transformed true residual noise via a strictly linear final activation:
\begin{equation}
    \hat{y}(\mathbf{r}) = \mathbf{W}_{\text{out}} * F_{\text{dec}}^{\text{final}} + \mathbf{b}_{\text{out}}
\end{equation}
By minimizing the Mean Squared Error, $\mathcal{L}_{\text{MSE}} = \frac{1}{N} \sum_{i=1}^{N} \| y_i - \hat{y}_i \|_2^2$, the network isolates the stochastic fluctuations and circumvents the mathematical complexity of reconstructing a high-dynamic-range physical signal from scratch.

\subsection{Deep Learning: Swin-Conv-UNET (SCUNET) Architecture}
As a state-of-the-art 2D comparative baseline, we evaluate the Swin-Conv-UNET (SCUNET)~\cite{Zhang2023}. SCUNET splits the input feature map to simultaneously apply Residual Convolutional (RConv) blocks ($Y_1$) for translation-invariant local features, and Swin Transformer (SwinT) blocks ($Y_2$) for non-local global self-attention. The parallel representations are fused and combined with the input via a residual connection:
\begin{equation}
    X_{\text{out}} = \text{Conv}_{1\times 1}(\text{Concat}(Y_1, Y_2)) + X_{\text{in}}
\end{equation}
This grouped split-and-fuse strategy maintains robust hierarchical feature extraction while significantly reducing the computational complexity of standard windowed attention.

\section{Results and Discussion}

\subsection{2D Image Models versus Volumetric Architectures}
\label{sec:image_results}
\begin{figure}[htbp]
    \centering
    \includegraphics[width=\linewidth]{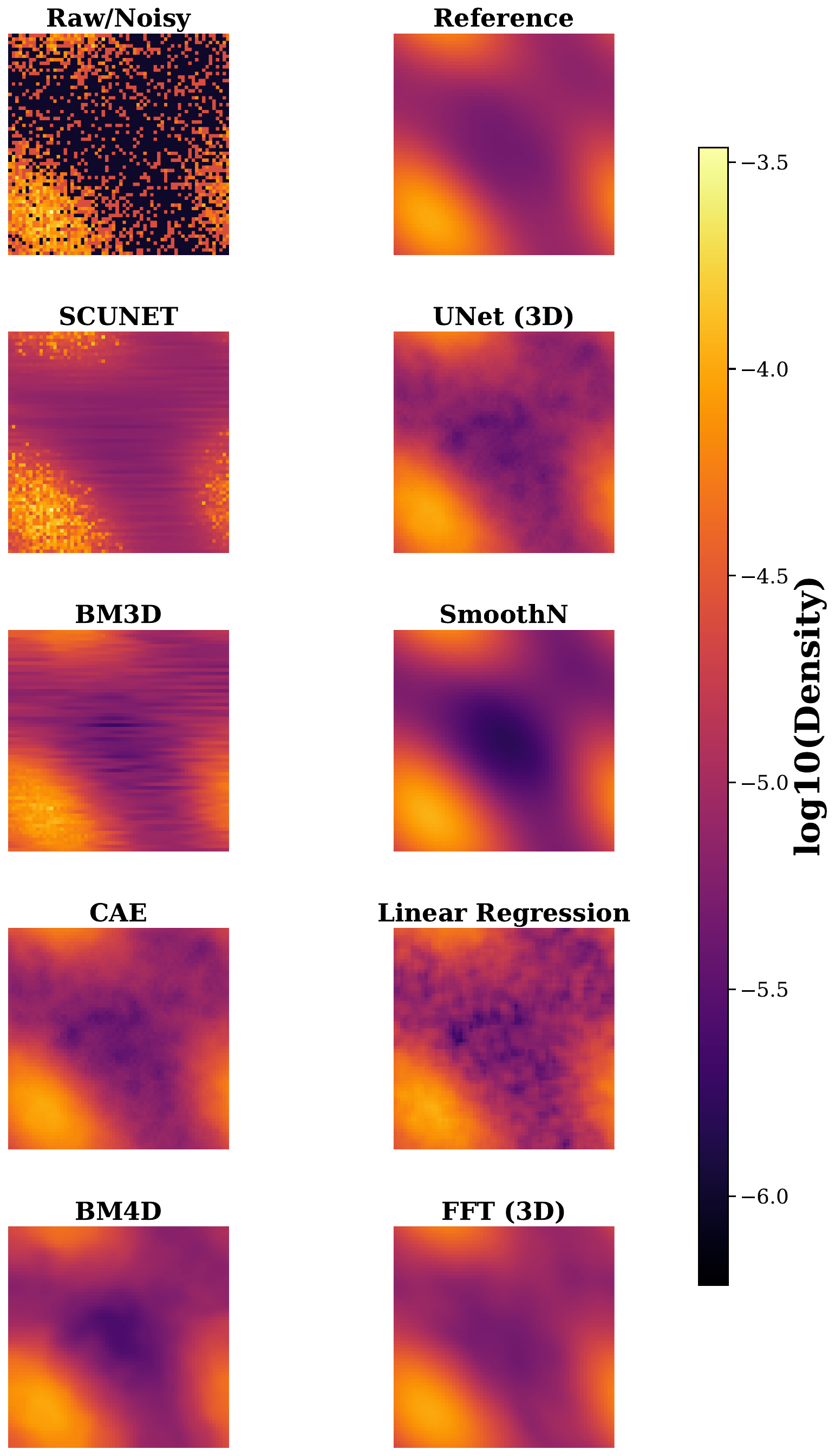}
    \caption{Visual comparison of a cross-sectional plane ($z=32$) of the diamond electron density profile denoised using various 1D, 2D image-based, and 3D volumetric methods. The comparison highlights structural anomalies, such as horizontal striping and low-density blurring, present in slice-by-slice 2D models compared to volumetric models.}
    \label{fig:3D_slice}
\end{figure}

To visualize the performance of different denoising models, we compare the denoising capabilities of off-the-shelf image denoisers and several of our volumetric denoisers on a VMC diamond electron density. We extracted a 2D cross-sectional slice ($z=32$) from each of the denoised densities for visual comparison, as illustrated in Figure~\ref{fig:3D_slice}. While the 1D sequential polynomial smoother (Linear) and 2D models like BM3D and SCUNET successfully recover the bulk electron distribution, they introduce some horizontal line artifacts and non-physical smearing into interstitial low-density regions due to their restricted spatial receptive fields. The standard 2D CAE model, trained slice-by-slice directly on the noise-injected DFT density matrices, does a great job at mitigating these artifacts relative to the pretrained SCUNET. Similarly, the 3D UNET model trained directly on the synthetic full volumetric data completely eliminates these slice-induced structural artifacts by natively accounting for three-dimensional correlations. Excellent qualitative agreement and smooth visual fidelity are likewise achieved by the FFT filter, the linear method, SmoothN, and the volumetric BM4D method.

\begin{figure}[htbp]
    \centering
    \includegraphics[width=\linewidth]{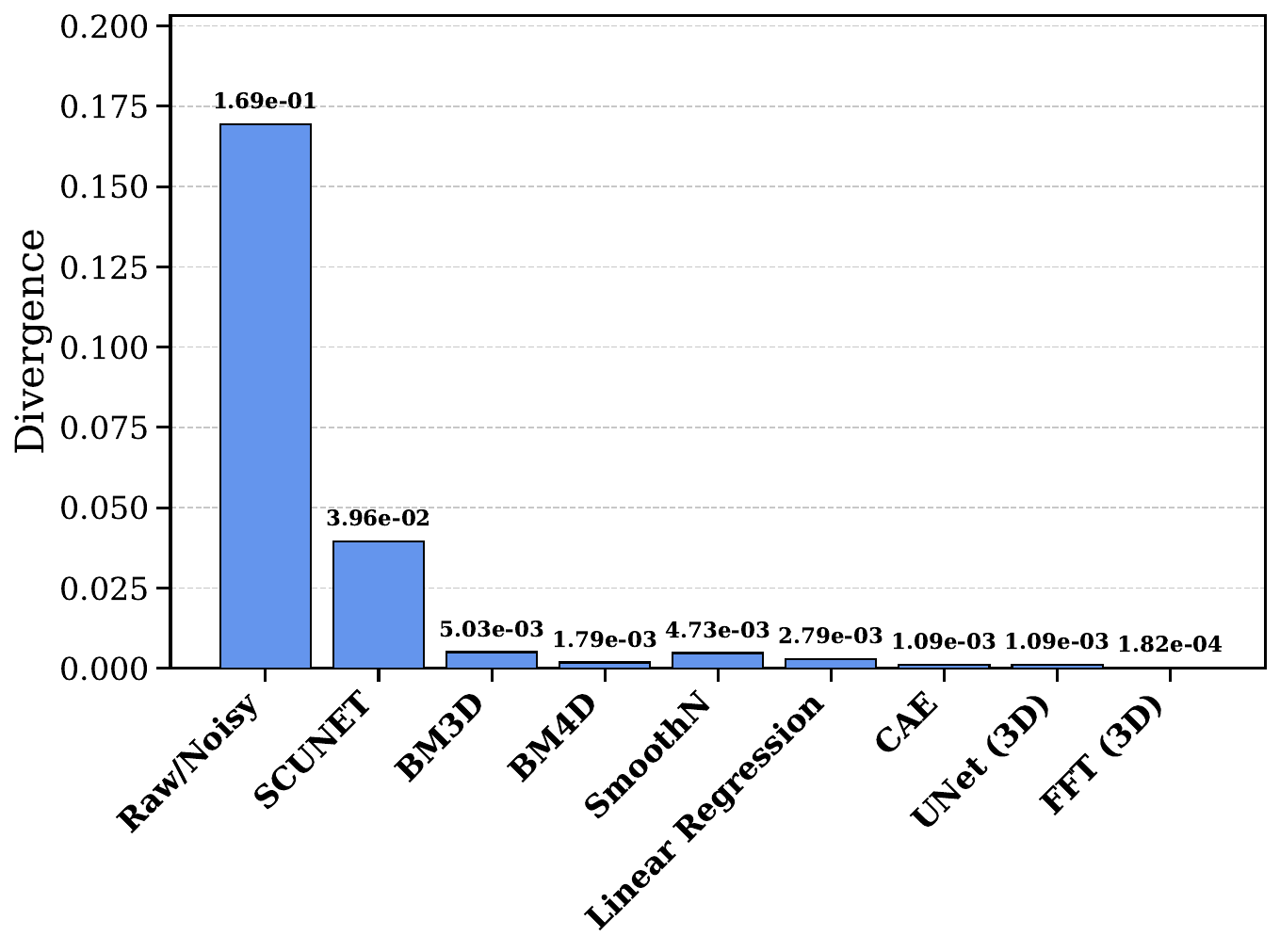}
    \caption{Direct JSD comparison for different image-based and volumetric denoising models evaluated against the diamond VMC reference density.}
    \label{fig:image_bar}
\end{figure}

Figure~\ref{fig:image_bar} provides a side-by-side bar chart comparison of the JSD values for each model. The primary takeaway from this perspective is twofold: first, our natively volumetric 3D models do not perform worse than established community image denoisers; second, standard 2D image denoisers are remarkably effective and highly applicable to volumetric denoising. Although the 3D UNET generates visually cleaner profiles with fewer non-physical discontinuities, the 2D models achieve highly competitive  JSD values. This indicates that the bulk of the statistical noise can be effectively mitigated even when artificially forcing a 2D matrix slicing. 

The viability of applying lower-dimensional models to volumetric data is further supported by the largely similar performance between the 2D BM3D and 3D BM4D algorithms across a broader sampling range (as detailed in the ``Estimating Sigma'' section of the supplementary material). Even more strikingly, the highly disadvantaged 1D smoother (Linear) performs similarly to complex 2D and 3D frameworks, as evidenced by its competitive JSD score in Figure \ref{fig:image_bar}. This underscores that, while native 3D cross-correlations are necessary for perfect structural fidelity, they are not strictly required to achieve a mathematically useful reduction in overall statistical divergence.

Nevertheless, while 2D image models are statistically viable, their direct application to electron density data still introduces certain architectural inefficiencies. First, off-the-shelf pre-trained weights for models like SCUNET \cite{Zhang2023} are optimized for synthetic camera sensor noise (e.g., AWGN, JPEG compression) rather than the heteroscedastic, Poisson-distributed shot noise characteristic of QMC. Furthermore, forcing a continuous 3D volumetric scalar field into a 2D network necessitates slicing the matrices into independent planes, a process that mathematically severs essential physical cross-correlations along the $z$-axis. 

Finally, natural images possess bounded, discrete pixel arrays (e.g., 8-bit integer values), whereas electron density maps exhibit extreme dynamic ranges---peaking sharply at nuclear coordinates and decaying exponentially into the vacuum \cite{Bader1990, Tal1978}. In contrast to these inherent 2D limitations, our natively trained 3D UNET model highlights the value of purpose-built, domain-specific architectures, effectively leveraging the foundational principle of a deep neural network bottleneck to isolate the core deterministic electron density signal from stochastic fluctuations \cite{Ronneberger2015, Cicek2016} while preserving the full 3D spatial topology.

\subsection{Role of Density Transforms and Asymptotic Divergence Corrections}

\begin{figure*}[!htbp]
    \centering
    \includegraphics[width=\linewidth]{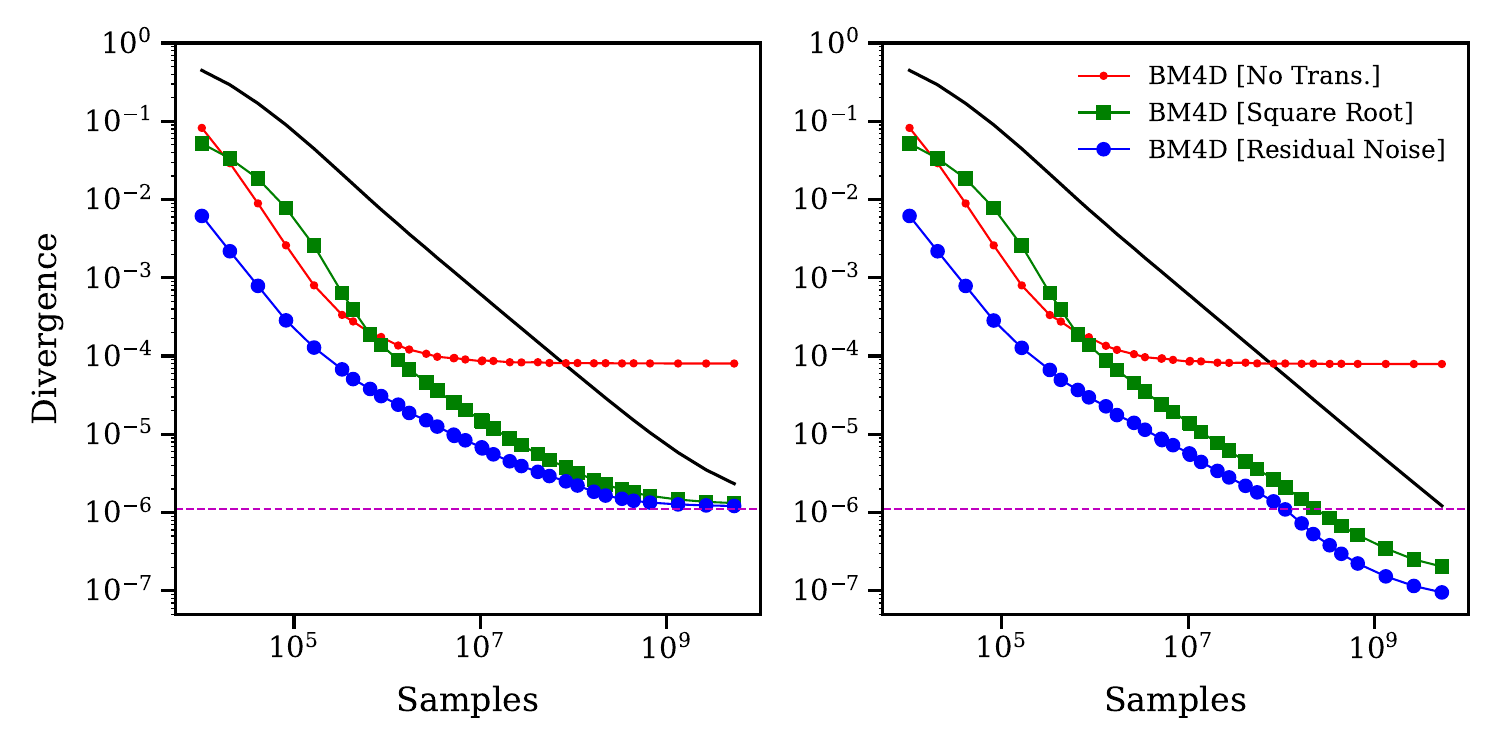}
    \caption{Divergence of the denoised densities generated using BM4D with different density transformations relative to the reference. The denoising quality improves as one progresses from no transform (``value'') to square root and residual signal regularized noise.  (Left) Bare divergences against the reference.  At high sampling, the apparent effectiveness of the denoising wanes (sqrt and residual noise transforms).  The horizontal dashed line demarcates the theoretical divergence floor when the high sampling reference is compared against the true noise-free density.  (Right) Similar to the left panel, but now with the asymptotic correction applied to approximate the divergence of a denoised sample with the true noise free reference.}
    \label{fig:trans_div_corr}
\end{figure*}

As alluded to earlier, transformations of the density can improve the performance of the denoising methods over the bare noisy data. The left panel of Figure~\ref{fig:trans_div_corr} illustrates the relative performance of the BM4D denoiser under different density transformations. A comprehensive comparison of these transformations across all evaluated methods is provided in Section~\ref{sec:trans_imp_all} of the Supplementary Information.  Each point on the plot corresponds to denoising a separate noisy density obtained with different levels of sampling effort, with the sampling/noise increasing/decreasing from left to right.  The black line is the divergence between the untouched noisy input densities and a reference density obtained at high sampling ($>$5 billion samples).  As can be seen from this log-log plot, the divergences of the bare noisy densities follow a power law, which is identical to the asymptotic form discussed in Sec. \ref{sec:Error_Qual}.  

The results of denoising with BM4D are shown as colored lines separately for each density transformation.  As can be seen, BM4D denoises effectively at low sampling/high-noise, but it rapidly plateaus, showing little to no further improvement beyond one million input samples. By contrast, the square root and residual transformations show significant improvement over most of the sampling range considered. In comparing these two transformations, the residual transform outperforms the square root transform at all sampling levels.    

The horizontal dashed line is the predicted fundamental limit in divergence when comparing the high sampling reference with the true noise-free limit.  As is evident, neither the denoisers nor the bare input density with highest sampling (equal to the reference) pass below this line. The close approach to this line by BM4D applied to transformed data suggests that substantial denoising is occurring, despite the otherwise apparent performance plateau.

The right panel in Fig.~\ref{fig:trans_div_corr} uses the asymptotic formulae in Sec.~\ref{sec:Error_Qual} to apply a correction to the divergence data, removing the effect of comparing against a reference with finite noise. 
As shown there, in reality, the performance of the denoisers remains uninterrupted even while processing densities generated with one billion samples.  All data presented below include this correction to improve the conclusions derived at very high sampling.

\subsection{Divergence and Speedup vs. Samples for Hyperparameter-Tuned Denoisers}

\begin{figure*}[htbp]
    \centering
    \includegraphics[width=\linewidth]{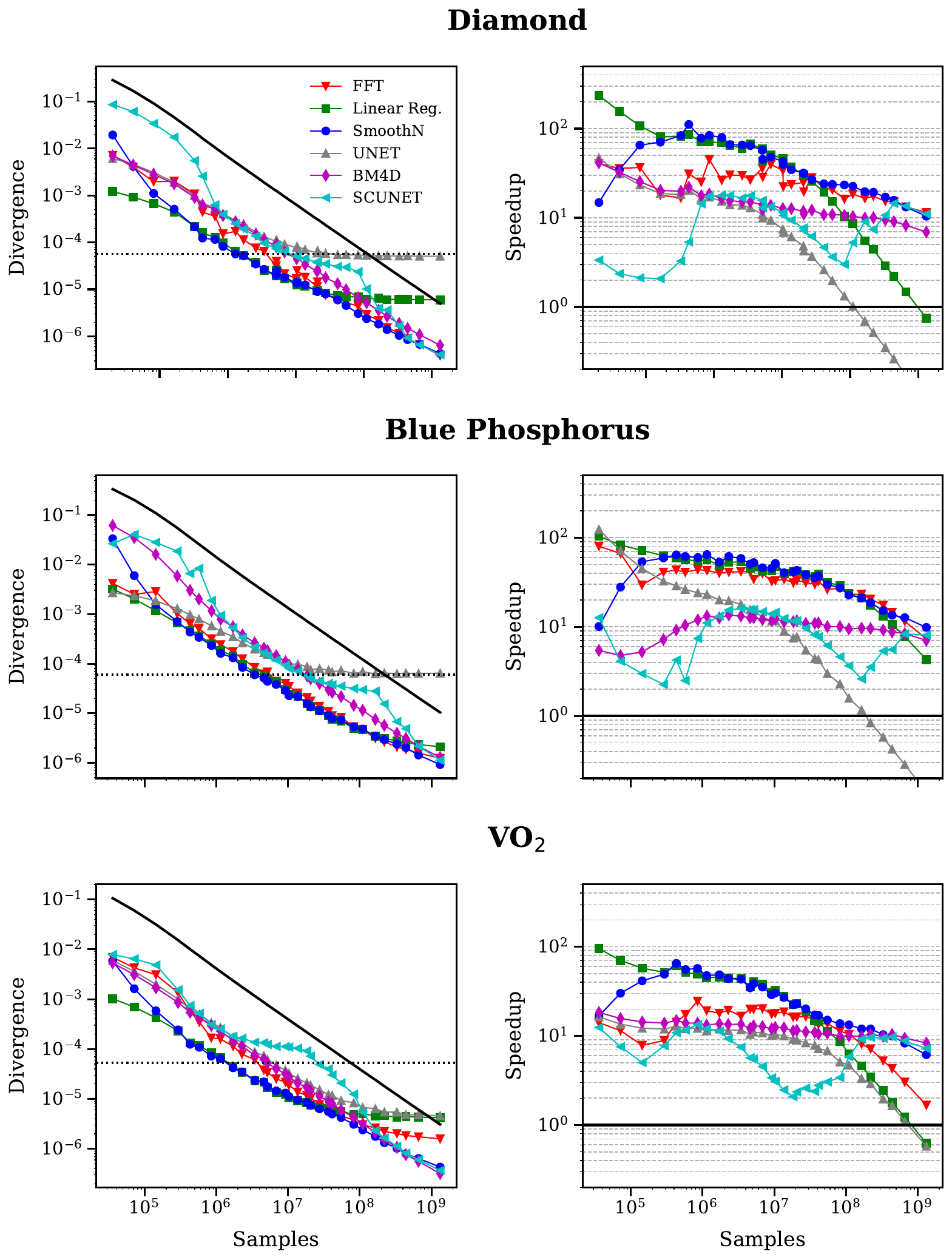}
    \caption{Divergences (left) and speedup (right) of denoised densities produced using hyper-tuned denoisers of each class on diamond (top), blue phosphorus (middle), and r-VO$_2$ (bottom). The horizontal dashed line is the divergence between the DFT density and the high-sampled reference, marking a threshold for densities achieving a ``useful'' noise reduction. Several denoisers display non-uniform noise reduction/increase across materials and input noise levels (sample counts). Multiple denoisers show distinctly different speedup behavior in the low and high sampling limits, with SmoothN and BM4D showing the greatest consistency.}
    \label{fig:div_samp_tuned_speedup}
\end{figure*}

By directly comparing the performance of the tuned denoising methods across materials, we gain insight into the noise regimes in which each method operates optimally. The left column of Figure~\ref{fig:div_samp_tuned_speedup} illustrates an improved $D_{JS}$ (y-axis) across all material systems compared to the baseline, utilizing a diverse suite of optimized denoisers (FFT, Local Regression, SmoothN, 3D UNET, BM4D, and SCUNET) where hyperparameters explicitly minimize the $D_{JS}$ at each discrete sampling level (x-axis). 
As a point of reference, we plot the JSD of the bare noisy densities computed against the high-sample reference DMC density as a solid black line.  The noisy data fall within the regime of validity for the low-noise asymptotic JSD formula, as evidenced by the near linearity of the JSD for the noisy densities.  

Broadly, two distinct classes of behavior are observed: some models excel in the low-sample/high-noise regimes, while others perform optimally in the high-sample/low-noise regimes. Local Regression, the FFT filter, and the 3D UNET all demonstrate robust performance under high-noise/low sampling conditions. Conversely, SCUNET and SmoothN exhibit superior efficacy in the low-noise/high-sampling regime. Both BM4D and FFT maintain a fairly uniform performance across varying sample counts. While BM4D seldom achieves the best performance for any individual sampling level, it is one of the few denoisers that provides effective noise reduction across all sample counts. Crucially, the practical utility of these denoisers is demarcated by comparison against the DFT baseline (the horizontal dotted line). 
%Any degree of raw sampling, or denoising, that does not exceed this threshold does not effectively reveal information unique to the higher accuracy DMC density.  
When a denoised density surpasses the DFT baseline while simultaneously improving upon the raw QMC divergence, it confirms that the filter is successfully exposing the inherently higher many-body accuracy of the QMC simulation without collapsing back to the baseline DFT representation.

One perspective on the benefit of denoising is that reduced QMC sampling is required to produce a density with an acceptable level of statistical bias. 
In this sense, denoising ``speeds up'' a DMC calculation, which can be quantified in terms of an effective number of samples corresponding to the reduced Jensen-Shannon divergence.
A general benefit of this representation of the impact of denoising is that it is expressed in a concrete manner, rather than the more abstract divergence value.

To quantify the computational value of these methods, we consider the low-noise asymptotic limit. Under this condition, these definitions provide:
\begin{align}
    D_{JS}(p_\mathrm{DN}   ,p_\mathrm{exact}) &= \frac{1}{8\log(2)}\frac{M}{N_\mathrm{eff}} \\
    D_{JS}(p_\mathrm{noisy},p_\mathrm{exact}) &= \frac{1}{8\log(2)}\frac{M}{N_\mathrm{mc}} \nonumber
\end{align}
We define the speedup $S$ as the ratio of the effective sample count to the input sample count required to reach equivalent statistical errors:
\begin{align}
    S &= \frac{N_\mathrm{eff}}{N_\mathrm{mc}} \\
      &= \frac{D_{JS}(p_\mathrm{noisy},p_\mathrm{exact})}{D_{JS}(p_\mathrm{DN},p_\mathrm{exact})} \nonumber \\
      &= \frac{D_{JS}(p_\mathrm{noisy},p_\mathrm{ref})-\frac{M}{8\log{2}N_\mathrm{ref}}}{D_{JS}(p_\mathrm{DN},p_\mathrm{ref})-\frac{M}{8\log{2}N_\mathrm{ref}}}
\end{align}
where the asymptotic correction in Eq.\ref{eq:div_corr} has been applied in order to directly use the computed divergence between the denoised density and low-noise reference (made from $N_\mathrm{ref}$ Monte Carlo samples).

The right column of Figure~\ref{fig:div_samp_tuned_speedup} illustrates how much compute time is saved in terms of sampling effort. At peak efficiency, these algorithms achieve up to a 100-fold speedup, with the majority consistently exceeding a 10-fold reduction in required sampling---representing substantial computational savings. Consistent with the divergence metrics, BM4D and the FFT filter display fairly flat and consistent speedup profiles across the evaluated range. In the low-to-medium noise regimes, Local Regression and SmoothN emerge as the most sample-efficient methods. However, in the highly stochastic (high-noise) regime, the efficacy of the 3D UNET and Local Regression methods drops significantly, while SCUNET joins the leading methods such as SmoothN. Apart from SCUNET, across most denoisers, a general decreasing trend in speedup is observed as the raw sample count increases; this decay reflects the fundamental difficulty of filtering in the extreme low-noise regime without introducing systematic bias.

\subsection{Conservative Denoiser Scoring and Reliability}

To robustly distinguish the performance of these methods, we define a performance score, which computes the speedup of each method relative to the maximum speedup achieved across all systems and methods. A highly valuable computational tool requires consistency across both varying materials and distinct noise regimes. In order to assess the different denoisers, we use the following procedure to define an overall ``worst-case'' performance score for each denoiser. First, for each material, compute the maximum speedup attained by any method---separately in the mid-sampling and high-sampling limits---and divide the speedup of each respective method by the maximum to form a per-material per-sampling range score.  To obtain the final score for each denoiser, take the minimum score across all materials and the two sampling levels. Thus, if a denoiser has a final score of 0.5, then it never performed worse than 50\% of the best denoiser across all contexts.
%we report the minimum score attained by each method within the medium-to-high noise limits as a fraction of the maximum score achieved by any method for . In this specific evaluation window, the QMC reference curve dips below the corresponding DFT-to-QMC baseline. 

\begin{figure}[htbp]
    \centering
    \includegraphics[width=\linewidth]{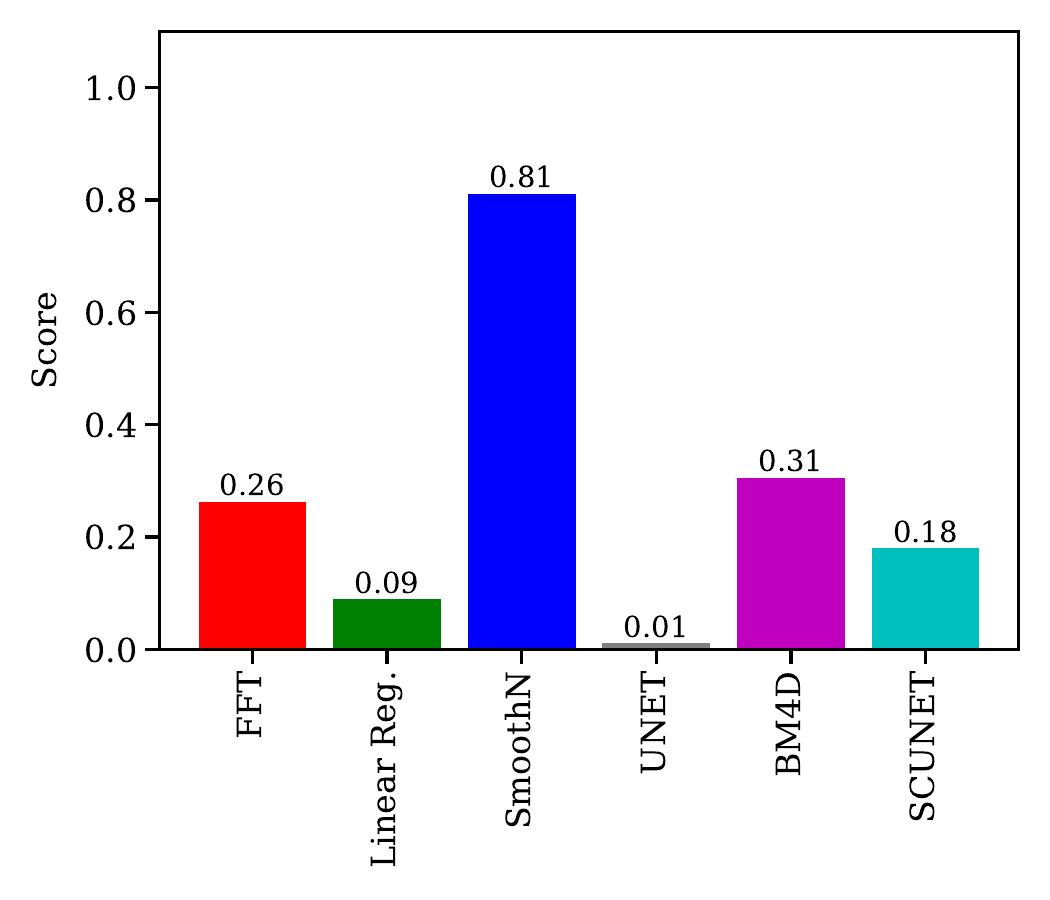}
    \caption{Performance scores computed for each denoiser. A score is the percentage of the maximum speedup across all denoisers attained by a particular denoiser. The performance score is the minimum score attained by a denoiser across materials in the mid and high sampling limits. It measures both the quality and the consistency of each denoiser. SmoothN shows both high consistency and performance.}
    \label{fig:cons_scores}
\end{figure}

Figure~\ref{fig:cons_scores} reveals that SmoothN provides the highest and most consistent performance across all systems, maintaining, at worst, 81\% of the maximum attainable speedup. This high worst-case threshold guarantees that selecting SmoothN \textit{a priori} heavily mitigates the risk of catastrophic algorithmic breakdown on unseen data. BM4D also demonstrates strong consistency, yielding the second-highest score at 31\%. The FFT filter is somewhat inconsistent across limits but scores favorably overall because lower-tier methods struggle significantly more in those same domains. Unsurprisingly, the simple linear local regression yields performance reflective of its basic mathematical formulation, securing a lower tier. Finally, considering the neural network-based approaches, both the trained 3D UNET and SCUNET architectures show poor generalizability under this rigorous minimum-score metric.

A clear performance hierarchy emerges based on the underlying mathematical machinery of the algorithms. The most robustly performant methods---BM4D \cite{maggioni2013} and SmoothN \cite{garcia2010}---both leverage the Discrete Cosine Transform (DCT), albeit through fundamentally different approaches. Garcia's SmoothN algorithm and BM4D diverge primarily in their spatial application of the DCT. SmoothN applies the DCT globally across the entire data grid to diagonalize a differential operator, effectively converting a massive, complex penalized least-squares linear system into a fast, point-by-point algebraic operation for spline-based smoothing \cite{garcia2010, garcia2011}, which is ultimately a regression approach. In contrast, BM4D applies the DCT locally and non-locally to clusters of highly correlated 3D image patches stacked into a 4D array, using the transform to sparsify signal components so that noise can be filtered via coefficient thresholding \cite{maggioni2013}, in a qualitatively similar manner to our FFT method. Ultimately, SmoothN utilizes the DCT as a mathematical accelerator to solve a global regression problem, whereas BM4D employs it as a statistical tool for transform-domain shrinkage of repetitive structural textures. Importantly, a lower score here does not strictly indicate overall model failure. Instead, it reflects variations in model reliability and generalization across distinct noise regimes and materials, which helps guide the optimal selection for practical denoising. A comprehensive comparison of speedups and scores for each method and material is provided in Section~\ref{sec:score_speed_all} of the Supplementary Information.

\subsection{Unified Cross-Material Scaling}

While many of the methods demonstrate exceptional performance well into the high-noise environments, our metrics (Figure~\ref{fig:cons_scores}) identified the regression-based SmoothN denoiser as the most statistically robust and generalized architecture across all evaluated sampling depths and material systems. Having established SmoothN as the optimal baseline, we conclude our analysis by interrogating its fundamental scaling behavior with respect to a unifying sampling metric: the number of electron visits per voxel. 

\begin{figure*}[htbp]
  \centering 
  \begin{subfigure}{0.48\textwidth}
    \centering
    \includegraphics[width=\linewidth]{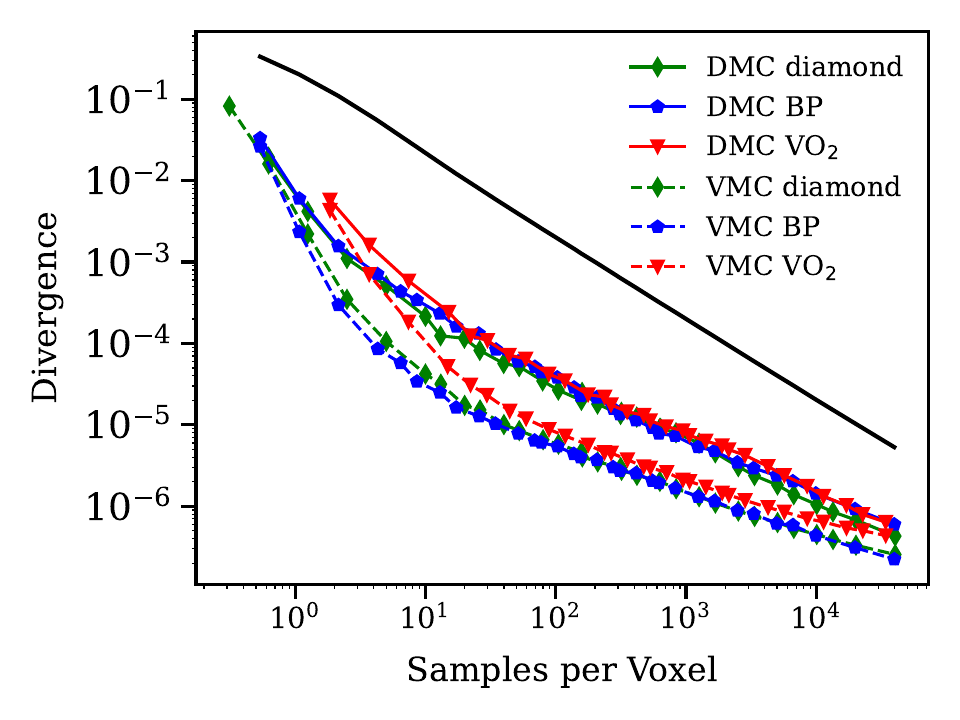}
    \label{fig:cross_mat_div}
  \end{subfigure}
  \hfill % Adds horizontal space to separate the images evenly
  \begin{subfigure}{0.48\textwidth}
    \centering
    \includegraphics[width=\linewidth]{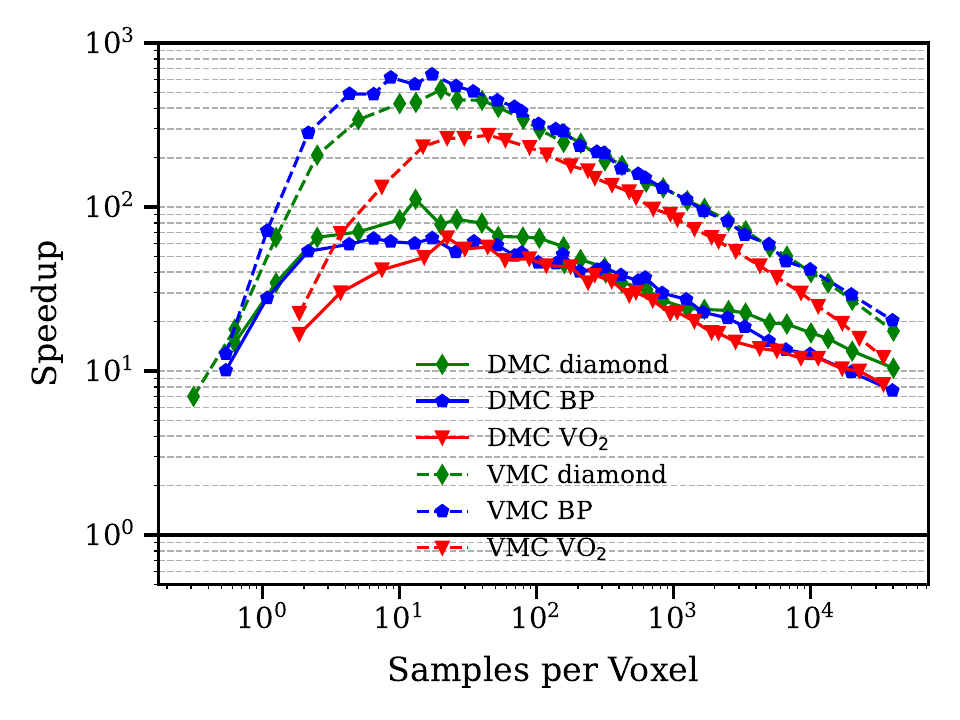}
    \label{fig:cross_mat_speedup}
  \end{subfigure}
  \caption{Performance of the SmoothN denoiser across all materials in terms of divergence (left) and speedup (right).  Alignment between the materials is achieved by plotting with respect to the number of samples per voxel (average samples per voxel) instead of the bare sample count.  The denoising results cluster tightly in a detailed way for both VMC and DMC.  At high number of samples per voxel, the speedup displays power-law behavior.}
  \label{fig:cross_mat_comp}
\end{figure*}

In Figure~\ref{fig:cross_mat_comp} (left), we observe a striking alignment in performance across each respective noise regime, demonstrating that the divergence tightly clusters based on the number of samples per voxel, despite the distinct morphological differences of the evaluated materials. Divergences for denoised VMC and DMC densities cluster around separate curves. When the the number of samples per voxel is low, the $D_{JS}$ is high, decreasing monotonically as the number of samples per voxel increases. Notably, SmoothN significantly suppresses the $D_{JS}$ for any given number of samples per voxel compared to the baseline stochastic QMC. 

The right panel of Figure~\ref{fig:cross_mat_comp} clarifies that, in terms of equivalent speedup, VMC data is generally more amenable to denoising than DMC data, yielding uniformly greater acceleration factors. Furthermore, as the number of samples per voxel increases, the speedup behavior exhibits an asymptotic decay that closely approximates a power law with sub-linear scaling, manifesting as a linear trend on a log-log scale. Extrapolating the trajectory of these linear log-log trends suggests that substantial computational benefits would persist even if the number of samples per voxel were increased by two orders of magnitude into the 500-billion sample regime, underscoring the extreme robustness and scalability of the SmoothN method.
This limit is truly extreme, where the task becomes finding vanishingly small noise without introducing systematic bias on the underlying smooth electron density.

%Overall, the top-performing methods across all regimes are deterministic filtering algorithms (SmoothN, FFT, and BM4D). The neural network models, while offering significant computational speedups in highly stochastic regimes, rank lower in overall robustness because they struggle to generalize across the entire divergence curve. However, given a restricted and well-defined noise regime, purely deep-learning models could likely be fine-tuned to achieve state-of-the-art, domain-specific performance, provided the reference bias observed at high-sample counts can be mitigated.

\section{CONCLUSION}

This study demonstrates that sophisticated post-processing denoising models offer orders-of-magnitude computational speedups in QMC electronic density generation compared to bare Monte Carlo sampling. Across a hierarchy of increasingly complex solid-state materials, we established that properly conditioned denoising algorithms can extract smooth, physically accurate electronic densities that closely mirror highly converged reference data. Crucially, the application of an approximate heteroscedastic to homoscedastic transformation proves essential for conditioning the density manifolds, enabling algorithms to isolate  and remove heteroscedastic QMC noise without corrupting the underlying deterministic signal.

While pre-trained 2D image models show good out-of-the-box performance, their 2D inductive biases and discrete grid formulations introduce unphysical slicing artifacts, highlighting the necessity for natively volumetric approaches. Among the tested methods, a 3D smoothing regression algorithm, SmoothN, exhibited the highest generalizability and consistency across all evaluated noise regimes and material morphologies. Although separate 3D neural networks comparatively struggle at the denoising task in the high-noise (SCUNET) and low-noise (UNET) limits, their strong performance on strongly correlated systems like VO$_{2}$ underscores their ability to capture complex spatial representations. Ultimately, purpose-built, volumetric denoisers represent a highly efficient avenue for QMC workflows, drastically lowering the sampling barrier required to utilize exact many-body densities in downstream electronic structure applications, such as the \textit{ab initio} development of density functionals.

While the volumetric denoising architectures evaluated in this study offer an immediate and powerful avenue for accelerating QMC workflows, several critical frontiers remain to be explored to maximize their physical fidelity, architectural generality, and downstream utility. A primary objective of future work is to deploy these models across a broader and more structurally diverse suite of chemical and material environments, encompassing low-dimensional interfaces, molecular crystals, and heterogeneous surfaces with strong correlation.

Simultaneously, the predictive capacity and representational power of our deep volumetric neural networks can be advanced through targeted architectural and data-engineering interventions. While the standard 3D UNET demonstrated strong speedups at intermediate sampling regimes, its convolutional layers are fundamentally constrained by local receptive fields. Integrating 3D self-attention mechanisms or axial attention blocks into the UNET bottleneck represents a major next step, enabling the network to dynamically weight and capture long-range, non-local multi-electron spatial correlations across the entire periodic simulation cell without exponentially compounding the parameter count.

Furthermore, because the performance of any supervised deep learning model is inherently tethered to its training distribution, significant focus must be directed toward refining our synthetic data generation protocols. Future frameworks will explore the generation of highly diverse synthetic densities that span a wider array of fictitious electronic configurations and pseudo-atomic coordinates. More importantly, we aim to implement target-conditioned data synthesis, wherein the synthetic training densities are explicitly engineered to match the local morphological, coordination symmetries, and spatial gradients of the specific physical system under investigation, thereby minimizing out-of-distribution errors during deployment.

Looking beyond architectural adjustments, future efforts will also aim to mitigate the underlying template bias that stems from our current reliance on baseline DFT reference grids for the variance transformation. While computationally inexpensive and highly effective for guiding spectral filters, this structural dependency risks suppressing exotic, multi-reference quantum fluctuations unique to many-body wavefunctions. Because inverse DFT procedures are historically hypersensitive to point-wise noise fluctuations, raw QMC densities present challenges in direct application to this process. In their current form, our stabilized, denoised densities represent computationally efficient, correlation-consistent QMC inputs for the extraction of high-quality Kohn-Sham exchange-correlation potentials directly from high-level many-body wavefunctions.

\section*{Acknowledgments}
%Work conducted bywriting) 

This work was primarily supported by the U.S. Department of Energy, Office of Science, Basic Energy Sciences, Materials Sciences and Engineering Division, as part of the Computational Materials Sciences Program and Center for Predictive Simulation of Functional Materials.  Authors  J.T.K. (concept, QMC runs, method/code development, analysis, writing, mentorship) and B.R. (mentorship, writing) acknowledge funding support from this source.
Work conducted by K.B. (method/code development, analysis, writing) was supported by the U. S. Department of Energy (DOE), Office of Science, Office of Workforce Development for Teachers and Scientists, Office of Science Graduate Student Research (SCGSR) program. The SCGSR program is administered by the Oak Ridge Institute for Science and Education (ORISE) for the DOE. ORISE is managed by Oak Ridge Associated Universities (ORAU) under contract number DE-SC0014664.

This research used resources of the National Energy Research Scientific Computing Center (NERSC), a U.S. Department of Energy Office of Science User Facility operated under Contract No. DE-AC02-05CH11231.
An award of computer time was provided by the Innovative and Novel Computational Impact on Theory and Experiment (INCITE) program. This research used resources of the Oak Ridge Leadership Computing Facility, which is a DOE Office of Science User Facility supported under Contract No. DE-AC05-00OR22725.

\section*{DATA AVAILABILITY}
The data supporting the findings of this study are available in the supplementary material 
with full data hosted by the Materials Data Facility\cite{Blaiszik2016} [link to be provided upon acceptance].

\bibliographystyle{naturemag}
\bibliography{Bib}
\end{document}

%\title{Supplementary Information for: Quantum Monte Carlo Electronic Density Denoising Methods}
\title{Supplementary Information for: Denoising Diffusion Monte Carlo Electron Densities with Physically Informed Variance Stabilization: From Fourier Filters to 3D UNETs}

\maketitle

\clearpage
\setcounter{figure}{0} % Resets the figure counter to 0
\renewcommand{\thefigure}{S\arabic{figure}} % Adds "S" before the number
\section{Full Details: Asymptotics of the Jensen-Shannon Divergence in the Low-Noise Limit}
\label{sec:asymp_deriv}
Here we consider the low-noise limit of the Jensen-Shannon divergence when 
comparing two noisy distributions/histograms with the same mean distribution.  
The form of the JS divergence is
\begin{align}
  D_{JS}(p_1,p_2) & = \frac{1}{2\log{2}}\sum_{m=1}^M\left[p_1(r_m)\log{\frac{p_1(r_m)}{p_{mix} (r_m)}}+p_2(r_m)\log{\frac{p_2(r_m)}{p_{mix}(r_m)}}\right]
\end{align}
where $p_{mix}$ is the mixture distribution $(p_1+p_2)/2$. 
A noisy histogram $\tilde{p}_m$ with cells indexed by $m$ can be written as
\begin{align}
    \tilde{p}_m = p_m + \delta p_m
\end{align}
where each $\delta p_m$ is a random variate with zero mean ($\mean{\delta p_m}=0$) and $p_m$ is the noise-free mean 
distribution/histogram.  In the small noise limit, we require each random variate to have infinitesimal variance, or
\begin{align}
    \textrm{var}(\delta p_m) &= \mean{\delta p_m^2}-\mean{\delta p_m}^2=\mean{\delta p_m^2}\ll 1
\end{align}

Our goal is to calculate the expected value of the JS divergence between two noisy distributions each having the same mean/noise-free distribution in the low-noise limit.
That is, we want $\mean{D_{JS}(\tilde{p}_1,\tilde{p}_2)}$ where $\tilde{p_1}=p+\delta p_1$ and $\tilde{p_2}=p+\delta p_2$.
Dropping the subscript $m$, we focus on the first logarithmic term in the JS divergence:
\begin{align}
    (p+\delta p_1)\log\frac{p+\delta p_1}{p+\tfrac{1}{2}(\delta p_1+\delta p_2)} &= (p+\delta p_1)\log\left(1+\frac{1}{2}\frac{\delta p_1-\delta p_2}{p+\tfrac{1}{2}(\delta p_1+\delta p_2)}\right)  \nonumber \\
    &\approx (p+\delta p_1)\log\left(1+\frac{\delta p_1-\delta p_2}{2p}\right) \nonumber
\end{align}
The full term in the brackets becomes
\begin{align}
    (p&+\delta p_1)\log\left(1+\frac{\delta p_1-\delta p_2}{2p}\right) + (p+\delta p_2)\log\left(1-\frac{\delta p_1-\delta p_2}{2p}\right)  \nonumber \\
    &= p\log\left(1-\frac{1}{4}\frac{(\delta p_1-\delta p_2)^2}{p^2}\right) + \delta p_1 \log\left(1+\frac{\delta p_1-\delta p_2}{2p}\right) + \delta p_2 \log\left(1-\frac{\delta p_1-\delta p_2}{2p}\right) \nonumber \\
    &\approx -\frac{1}{4}\frac{(\delta p_1-\delta p_2)^2}{p}+\frac{1}{2}\frac{\delta p_1^2-2\delta p_1\delta p_2+\delta p_2^2}{p} \nonumber \\
    &= \frac{1}{4}\frac{\delta p_1^2-2\delta p_1\delta p_2+\delta p_2^2}{p} \nonumber
\end{align}
Its expectation value is 
\begin{align}
\frac{1}{4}\left\langle\frac{\delta p_1^2-2\delta p_1\delta p_2+\delta p_2^2}{p}\right\rangle &= \frac{1}{4}\frac{\mean{\delta p_1^2}+\mean{\delta p_2^2}}{p}
\end{align}
Therefore, the expectation value of the divergence in the low-noise limit is
\begin{align}
    \mean{D_{J2}(\tilde{p_1},\tilde{p}_2)} &= \frac{1}{8\log 2}\sum_{m=1}^M \frac{\mean{\delta p_{1m}^2}+\mean{\delta p_{2m}^2}}{p}
\end{align}

For our particular case of electron density histograms, the noise in each cell follows a Gaussian/normal distribution with zero mean and a variance of $\textrm{var}(\delta p_m)=\mean{\delta p_m^2}=p_m/N$, where $N$ is the total number of samples drawn from $p$ over all histogram cells.  The expected divergence is then
\begin{align}
    \mean{D_{J2}(\tilde{p_1},\tilde{p}_2)} &= \frac{1}{8\log 2}\sum_{m=1}^M\left(\frac{1}{N_1}+\frac{1}{N_2}\right) \nonumber \\
    &=  \frac{1}{8\log 2}\left(\frac{M}{N_1}+\frac{M}{N_2}\right)
\end{align}
It follows that the divergence between a noisy distribution and its noise-free mean is 
\begin{align}
    \mean{D_{J2}(\tilde{p},p)} &= \frac{1}{8\log 2}\frac{M}{N}
\end{align}
%%%%%%%%%%%%%%%%%%%%%%%%%%%
\section{Estimating Sigma}\label{sec:est_sig}
%%%%%%%%%%%%%%%%%%%%%%%%%%%
In QMC simulations, the electron density is derived from stochastic sampling. Because the local sample counts ($N$) are often insufficient to satisfy the asymptotic limits of the Central Limit Theorem, the noise profile remains strictly signal-dependent, approximating a Poisson distribution with variance $\text{Var}(d_{\text{QMC}}) \approx d_{\text{QMC}}/N$. This signal dependence severely violates the Additive White Gaussian Noise (AWGN) assumption inherent in standard spatial denoisers such as BM3D and wavelet-based noise estimators. To bridge this statistical disparity, we employ a Variance Stabilizing Transformation (VST) defined as $d_{\text{trans}} = (d_{\text{QMC}} - d_{\text{DFT}}) / \sqrt{d_{\text{DFT}}}$, utilizing a baseline Density Functional Theory mean, $d_{\text{DFT}}$. Applying the variance operator to this transformation yields $\text{Var}(d_{\text{trans}}) = \frac{1}{N} (d_{\text{QMC}} / d_{\text{DFT}})$. While Variational and Diffusion Monte Carlo (VMC/DMC) densities capture many-body correlations that systematically differ from DFT, the baseline DFT density serves as a sufficiently accurate structural proxy such that $d_{\text{QMC}} / d_{\text{DFT}} \approx 1$. Consequently, the VST effectively decouples the noise variance from the extreme dynamic range of the local molecular topology, mapping it to a nearly uniform standard deviation of $\sigma \approx 1/\sqrt{N}$. The mathematical efficacy of this stabilization is empirically corroborated by evaluating the transformed matrices with a wavelet-based noise estimator (\texttt{skimage.restoration.estimate\_sigma}); the estimator successfully isolates the high-frequency spatial jitter to reliably reproduce the analytically anticipated $\sigma \approx 1/\sqrt{N}$, confirming the validity of applying Gaussian-centric denoising models to low-count QMC density data despite the minor local variance deviations driven by the $d_{\text{QMC}} / d_{\text{DFT}}$ structural ratio.

\begin{figure}[htbp]
    \centering
    \includegraphics[width=\linewidth]{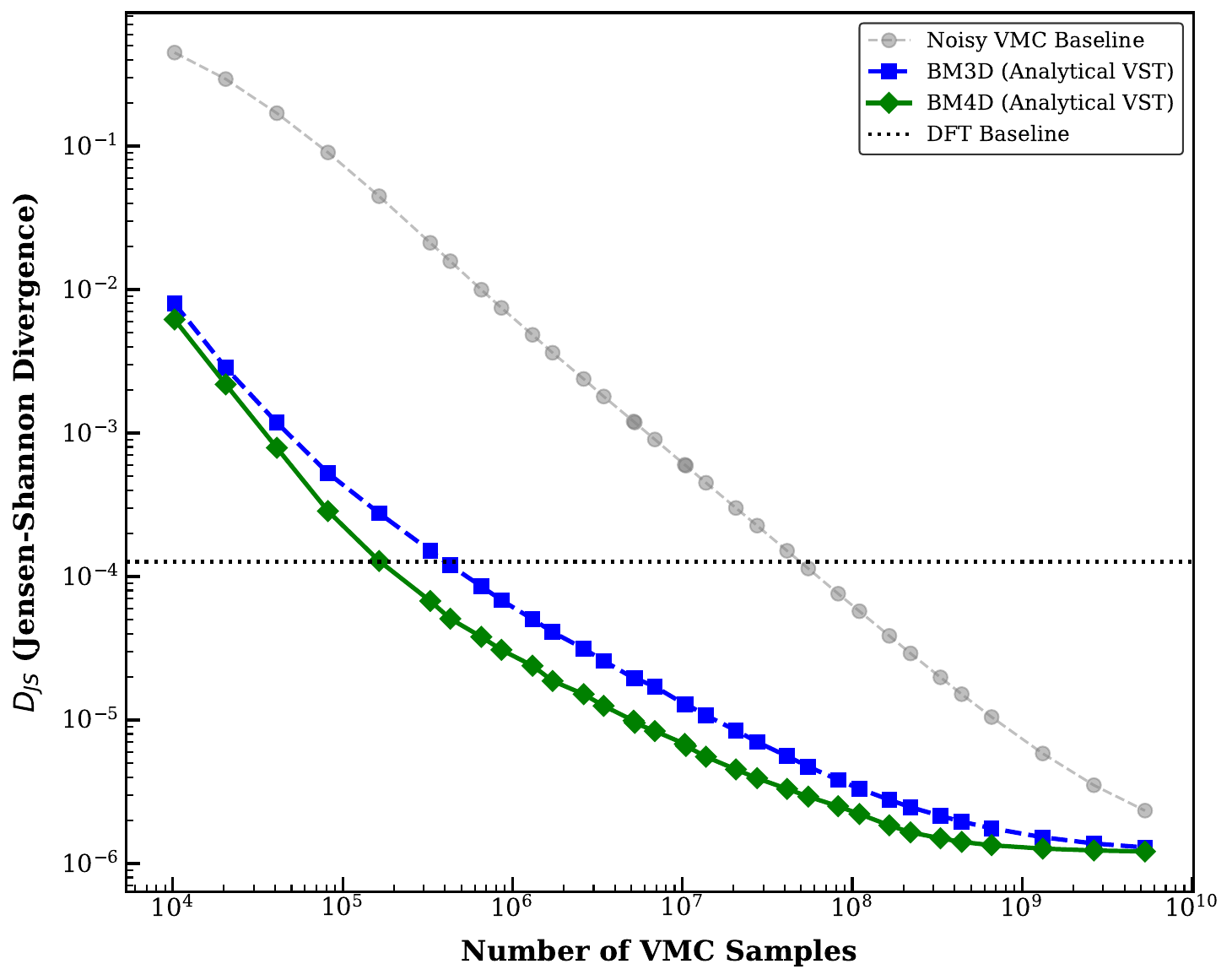}
    \caption{JSD of BM4D and BM3D using $\sigma=1/\sqrt{N}$}
    \label{fig:trans_sample}
\end{figure}

\section{Impact of Transformations on Denoiser Performance}\label{sec:trans_imp_all}

To systematically determine the overall impact of data transformation on denoising efficacy, we compare the performance of each denoiser across all materials using the available variance-stabilizing transformations. In addition to the standard transformations discussed in the main text, we evaluate several method-specific algorithmic variants where applicable. For the Fast Fourier Transform (FFT), alongside the standard filter, we evaluate a zero-value variant (denoted as \texttt{fftz\_val}), in which the high-frequency Fourier components are strictly zeroed out rather than being replaced by their corresponding density functional theory (DFT) spectral amplitudes. For local linear regression, we compare three variations of the sequential 1D polynomial smoother. The baseline \texttt{linear} model employs untuned hyperparameters, whereas \texttt{linear2} utilizes fully tuned hyperparameters. A third variant, \texttt{linear3}, flattens the native 3D density matrix into a 1D array and performs a simple ordinary least squares regression to fit a second-degree polynomial over a highly localized moving window of four neighboring points, strictly without enforcing periodic boundary conditions. Finally, the \texttt{SmoothN} label corresponds to the algorithm operating with untuned hyperparameters, while \texttt{SmoothN2} indicates the use of explicitly tuned hyperparameters. The optimal choice of transformation exhibits notable method dependence. For example, when employing the SCUNET architecture, the square-root transformation proves most effective in the low-sample (high-noise) regime, whereas the residual transformation performs best in the high-sample (low-noise) regime across all evaluated materials. Conversely, for the SmoothN algorithm, the residual transformation consistently yields the greatest denoising improvement across all sampling regimes for every material tested.

\begin{figure*}[htbp]
    \centering
    \includegraphics[width=\linewidth]{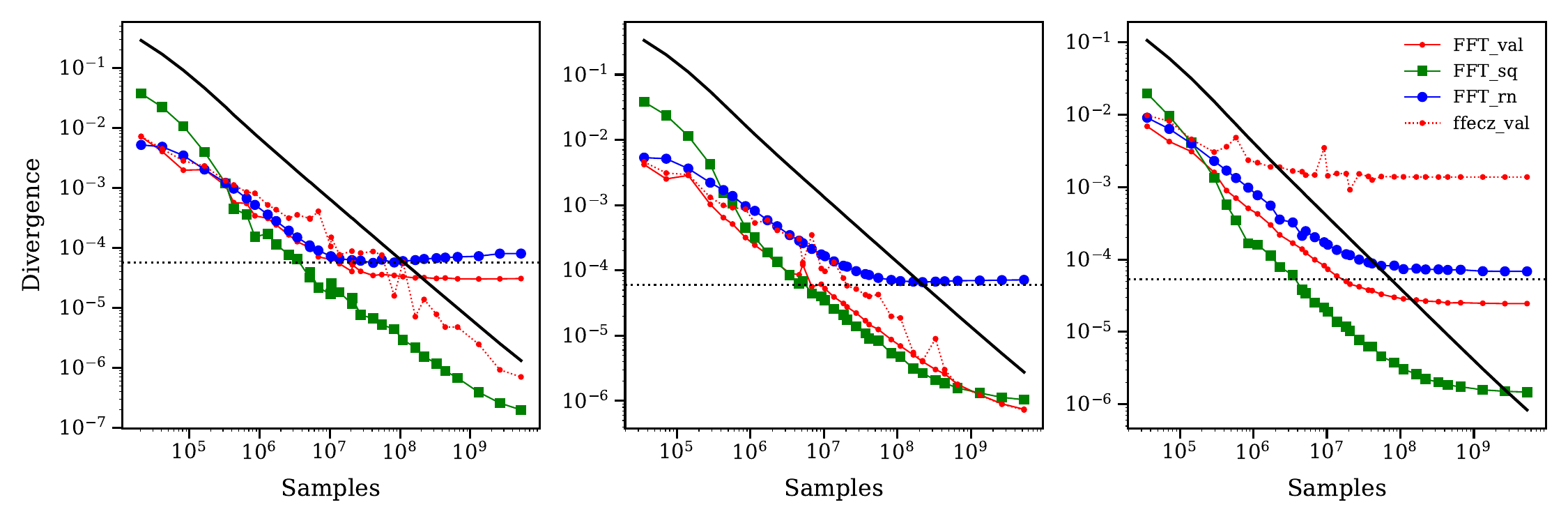}
    \caption{Divergence of DMC electronic densities for diamond, blue phosphorus, and VO$_2$ denoised using FFT across all transformations.}
    \label{fig:trans_div_FFT}
\end{figure*}

\begin{figure*}[htbp]
    \centering
    \includegraphics[width=\linewidth]{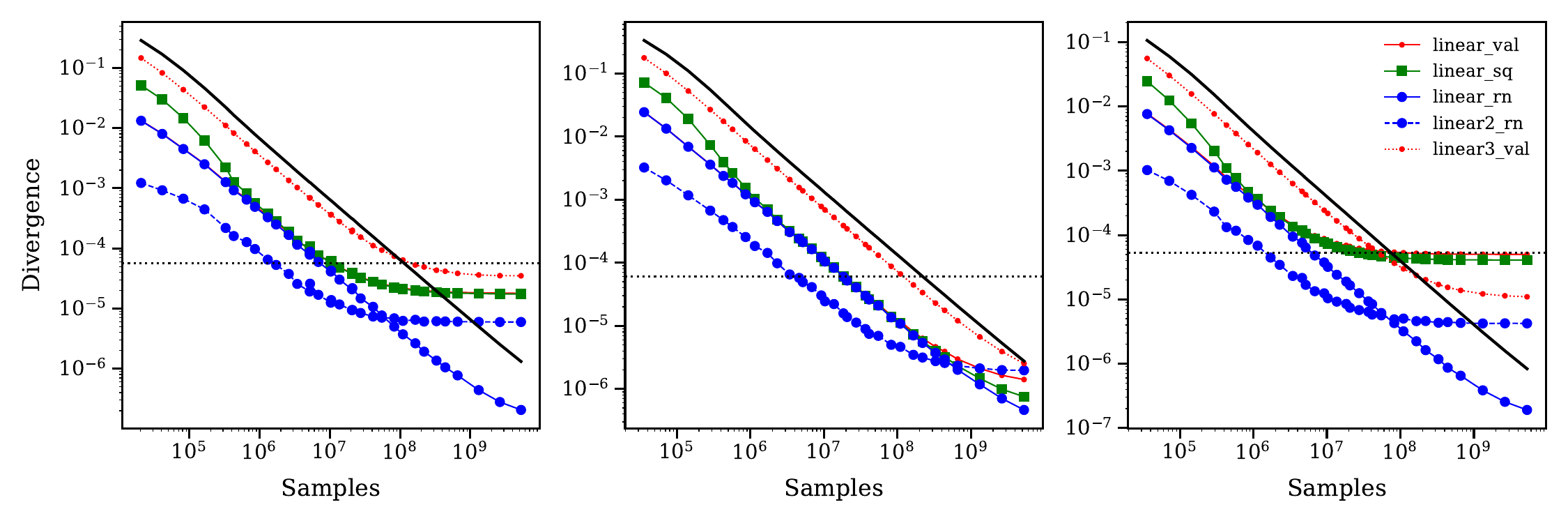}
    \caption{Divergence of DMC electronic densities for diamond, blue phosphorus, and VO$_2$ denoised using Linear Regression across all transformations.}
    \label{fig:trans_div_lin}
\end{figure*}

\begin{figure*}[htbp]
    \centering
    \includegraphics[width=\linewidth]{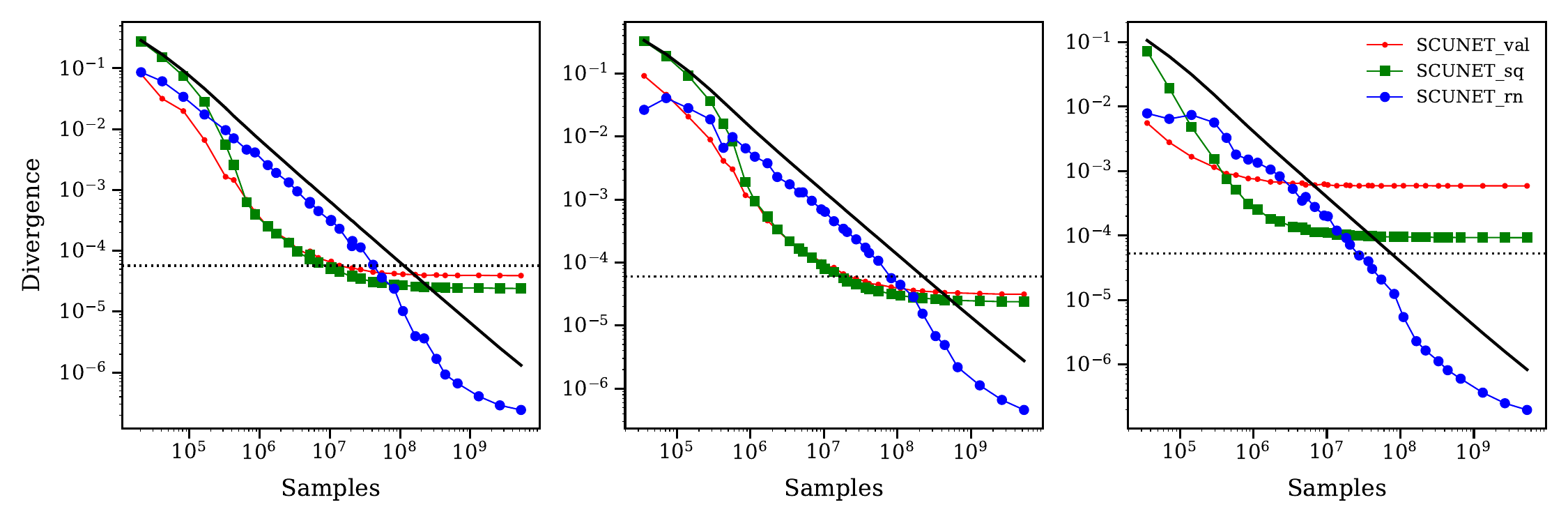}
    \caption{Divergence of DMC electronic densities for diamond, blue phosphorus, and VO$_2$ denoised using SCUNET across all transformations.}
    \label{fig:trans_div_SCUNET}
\end{figure*}

\begin{figure*}[htbp]
    \centering
    \includegraphics[width=\linewidth]{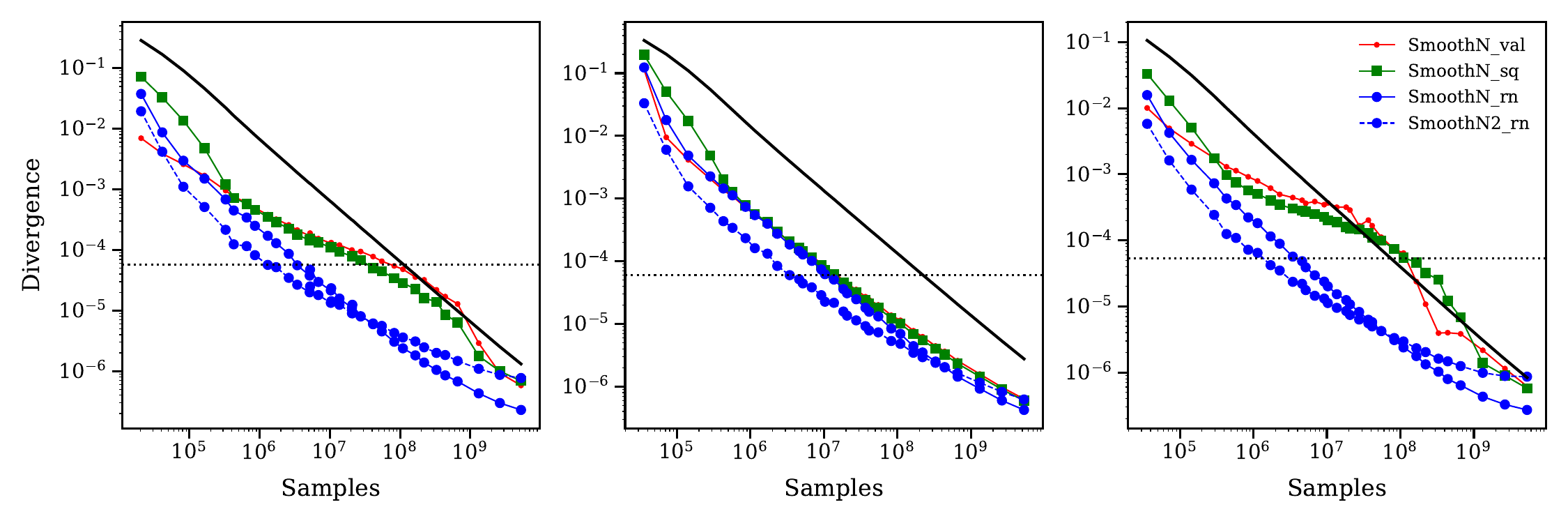}
    \caption{Divergence of DMC electronic densities for diamond, blue phosphorus, and VO$_2$ denoised using SmoothN across all transformations.}
    \label{fig:trans_div_SN}
\end{figure*}

\begin{figure*}[htbp]
    \centering
    \includegraphics[width=\linewidth]{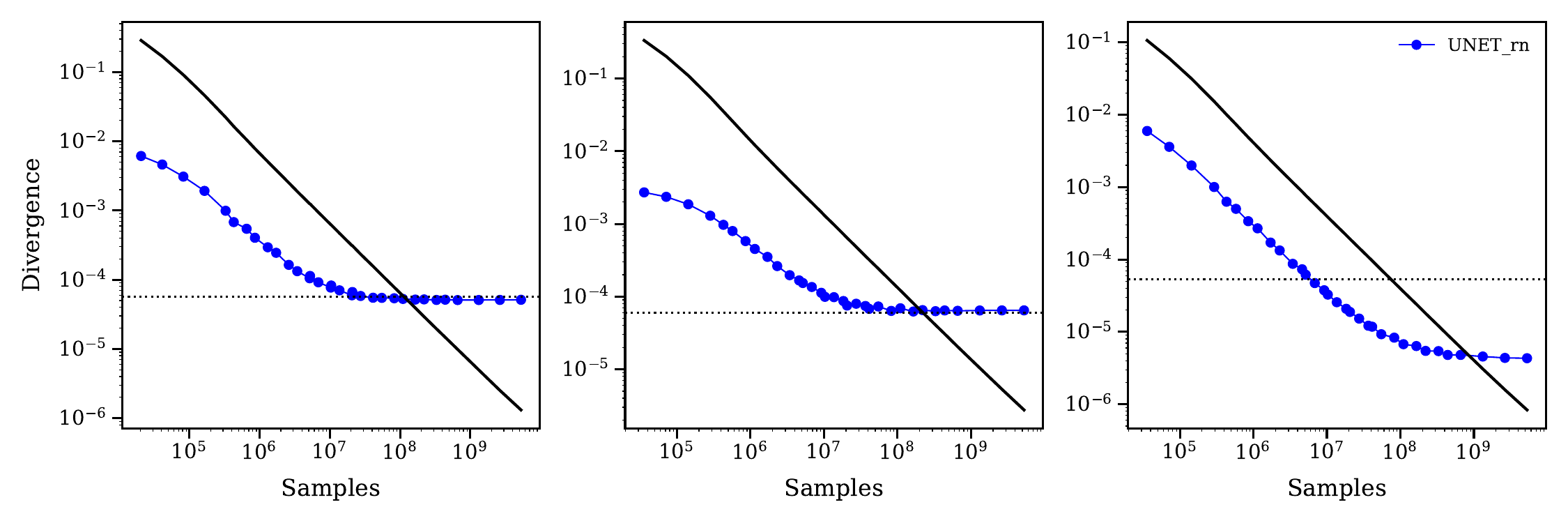}
    \caption{Divergence of DMC electronic densities for diamond, blue phosphorus, and VO$_2$ denoised using UNET across all transformations.}
    \label{fig:trans_div_UNET}
\end{figure*}

\begin{figure*}[htbp]
    \centering
    \includegraphics[width=\linewidth]{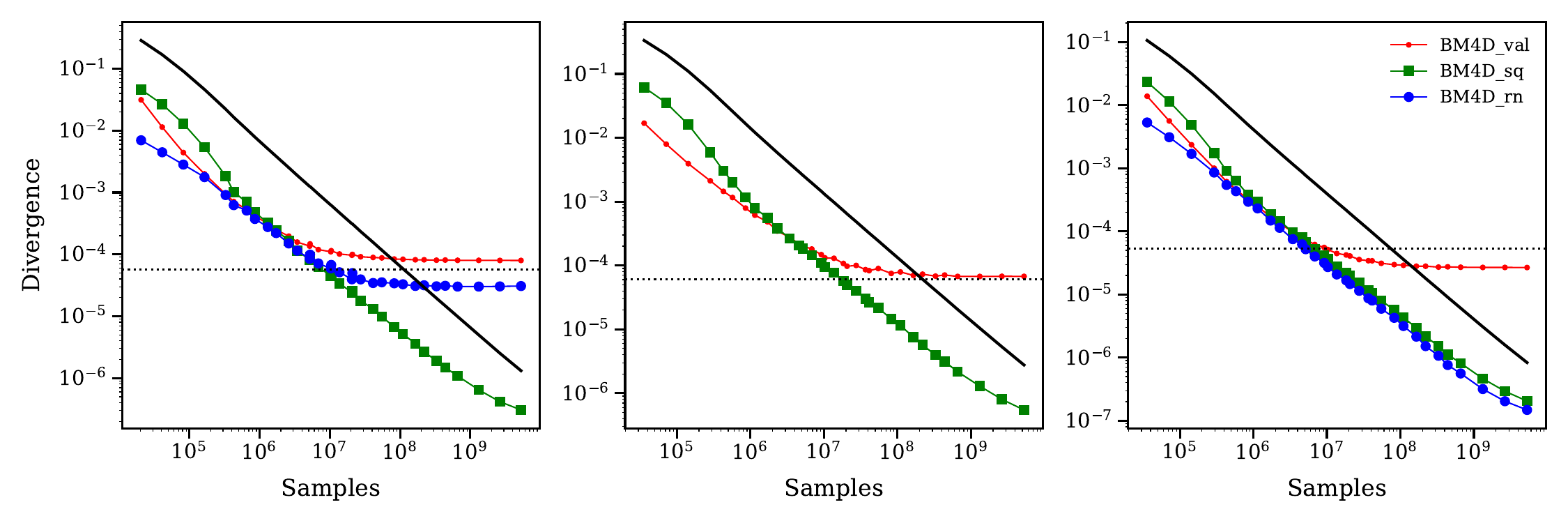}
    \caption{Divergence of DMC electronic densities for diamond, blue phosphorus, and VO$_2$ denoised using BM4D across all transformations.}
    \label{fig:trans_div_BM4D}
\end{figure*}

\section{Speedups and Scores by Sampling Level}\label{sec:score_speed_all}

We evaluate the computational speedups and relative scores for each material system across the medium and high sampling regimes. Transitioning to higher sampling levels generally results in reduced absolute speedups and relative scores. Furthermore, denoiser consistency varies significantly depending on the material and sample count; while some methods perform robustly across all test cases, others are only effective under specific conditions. Finally, analyzing the minimum relative scores provides insight into the worst-case performance of each denoising method across the tested materials for a given sampling level.

\begin{figure*}[htbp]
    \centering
    \includegraphics[width=\linewidth]{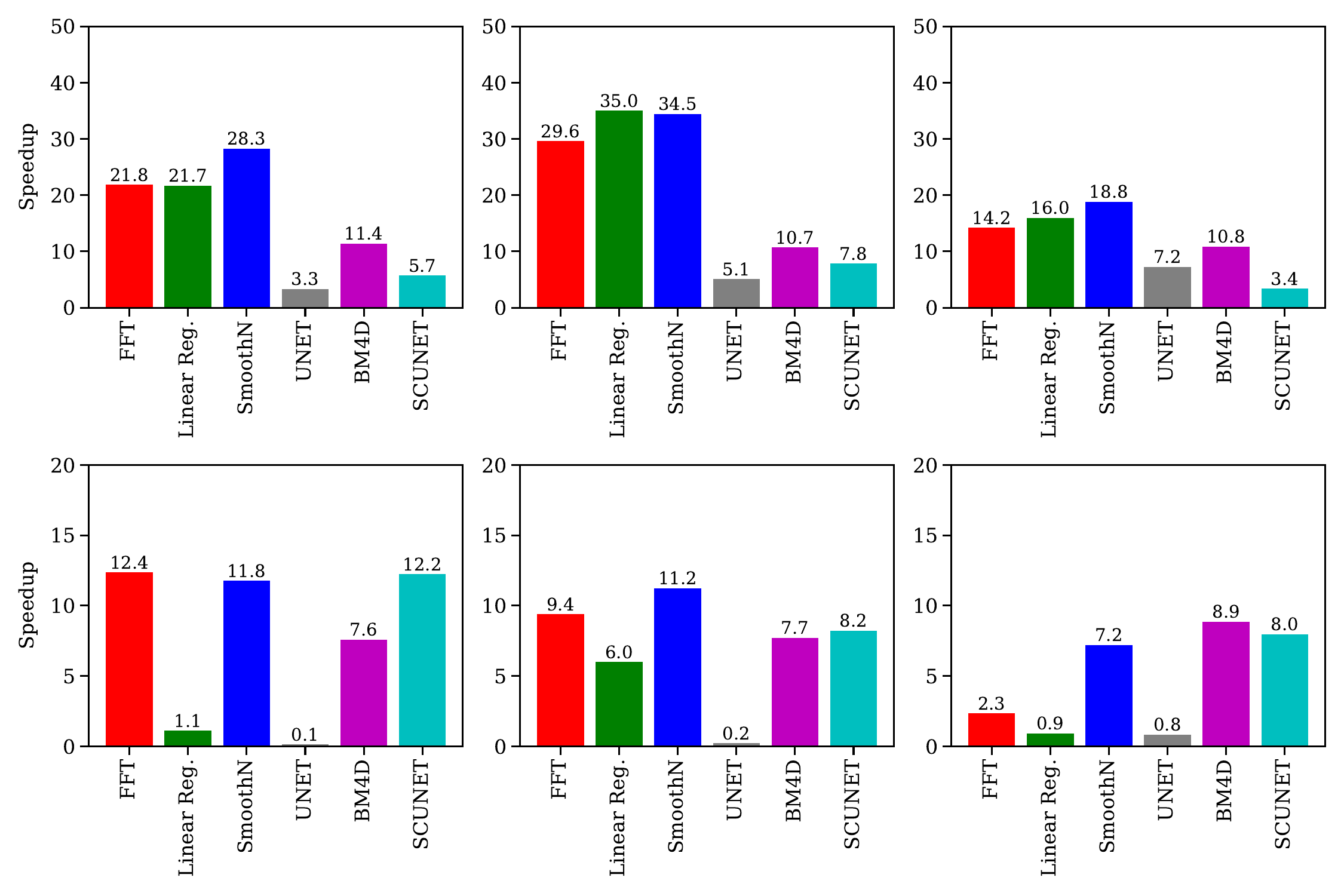}
    \caption{Speedup values across denoising methods for Diffusion Monte Carlo (DMC). The columns show data for diamond (left), blue phosphorus (middle), and VO$_2$ (right). The rows correspond to the medium (top) and high (bottom) sample regimes.}
    \label{fig:speedup_samplev_dmc}
\end{figure*}

%%%%%%%%%%%%%%%%%%%%%%%%%%%%%%%%%%%%%%%%%%%%%%%%%%%%%%%

\begin{figure*}[htbp]
    \centering
    \includegraphics[width=\linewidth]{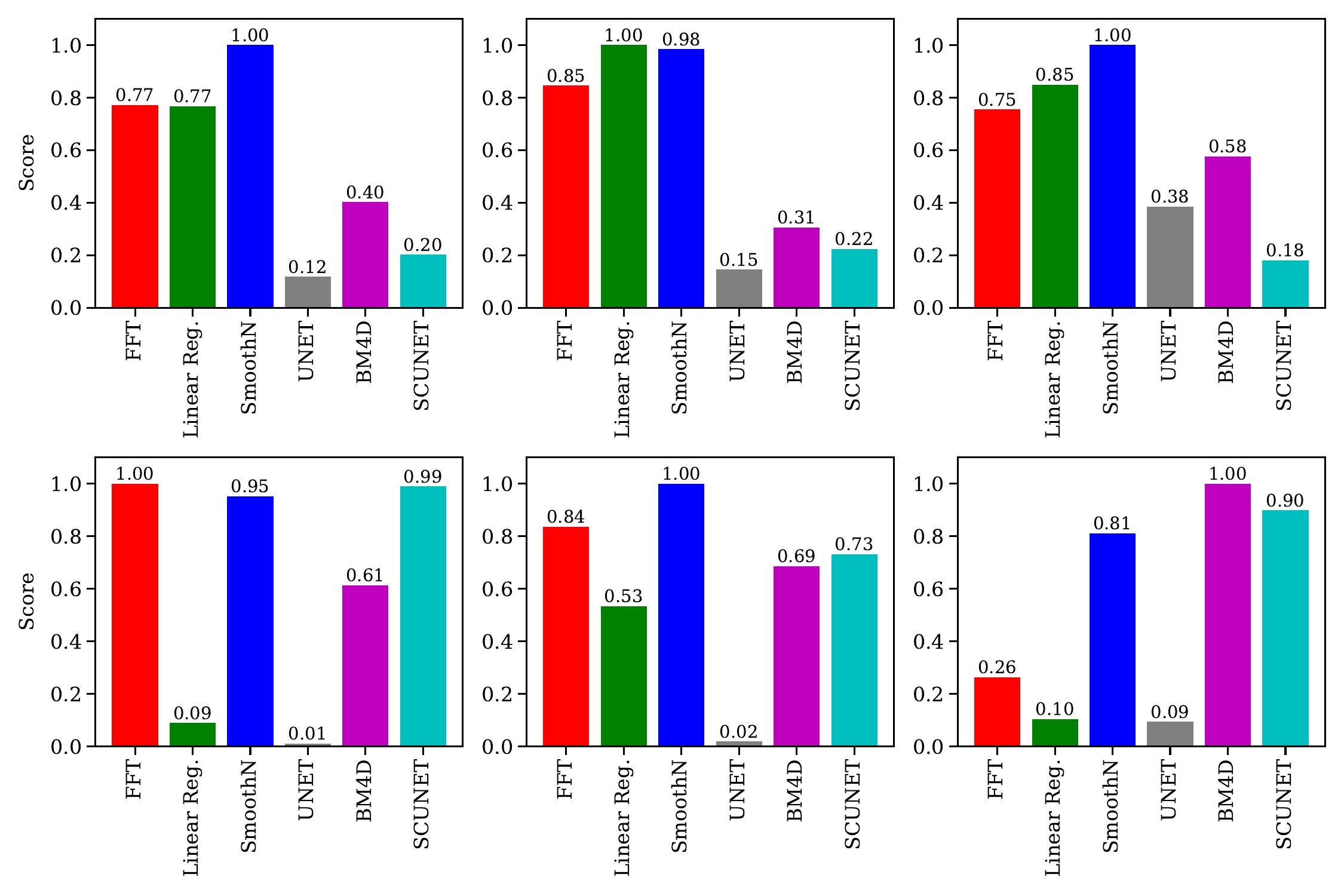}
    \caption{Relative speedup scores for each denoising method using Diffusion Monte Carlo (DMC). The columns correspond to diamond (left), blue phosphorus (middle), and VO$_2$ (right). The rows show results for the medium (top) and high (bottom) sample regimes.}
    \label{fig:score_samplev_dmc}
\end{figure*}

%%%%%%%%%%%%%%%%%%%%%%%%%%%%%%%%%%%%%%%%%%%%%%%%%%%%%%%

\begin{figure*}[htbp]
    \centering
    \includegraphics[width=\linewidth]{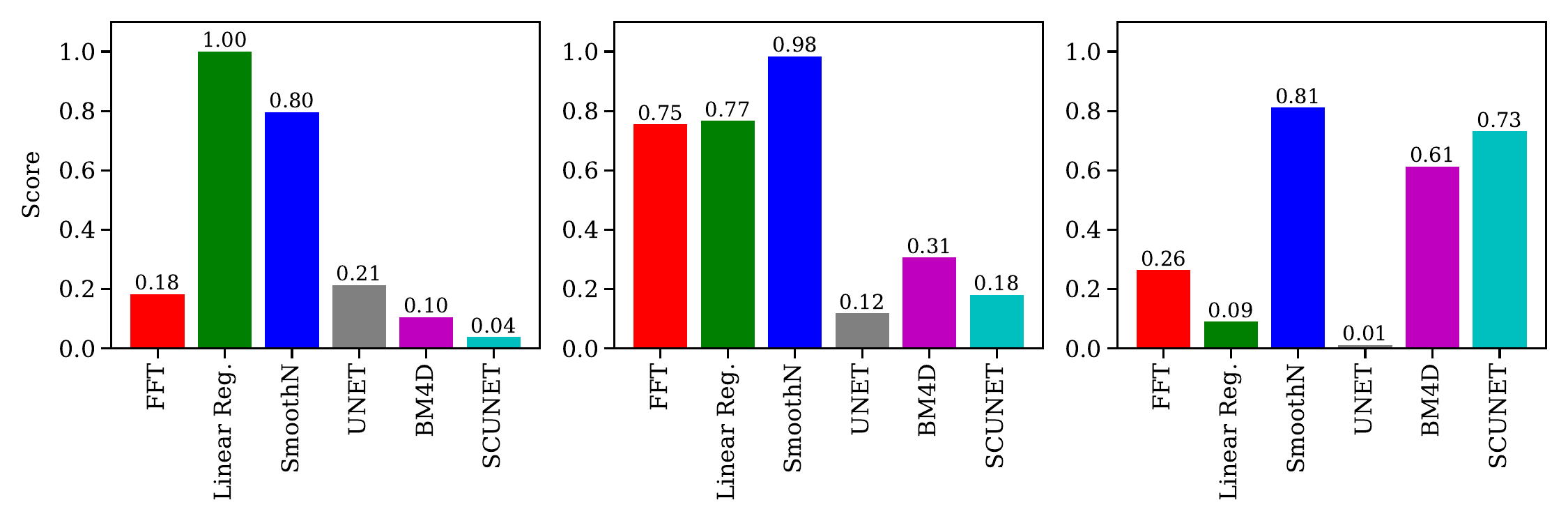}
    \caption{Minimum relative speedup score obtained by each denoising method across all three materials (diamond, blue phosphorus, VO$_2$) using Diffusion Monte Carlo (DMC). The panels correspond to the low (left), medium (middle), and high (right) sample regimes.}
    \label{fig:score_samplev_min_dmc}
\end{figure*}
%%%%%%%%%%%%%%%%%%%%%%%%%%%%%%%%%%%%%%%%%%%%%%%%%%%%%%%

\section{Visual Bias Check}\label{sec:Bias_check}
\par To determine whether our UNET model introduces bias into our electronic densities, we denoised several samples---including the reference density---to check for visual artifacts on VO$_2$.\begin{figure}[htbp]
    \centering
    \includegraphics[width=\linewidth]{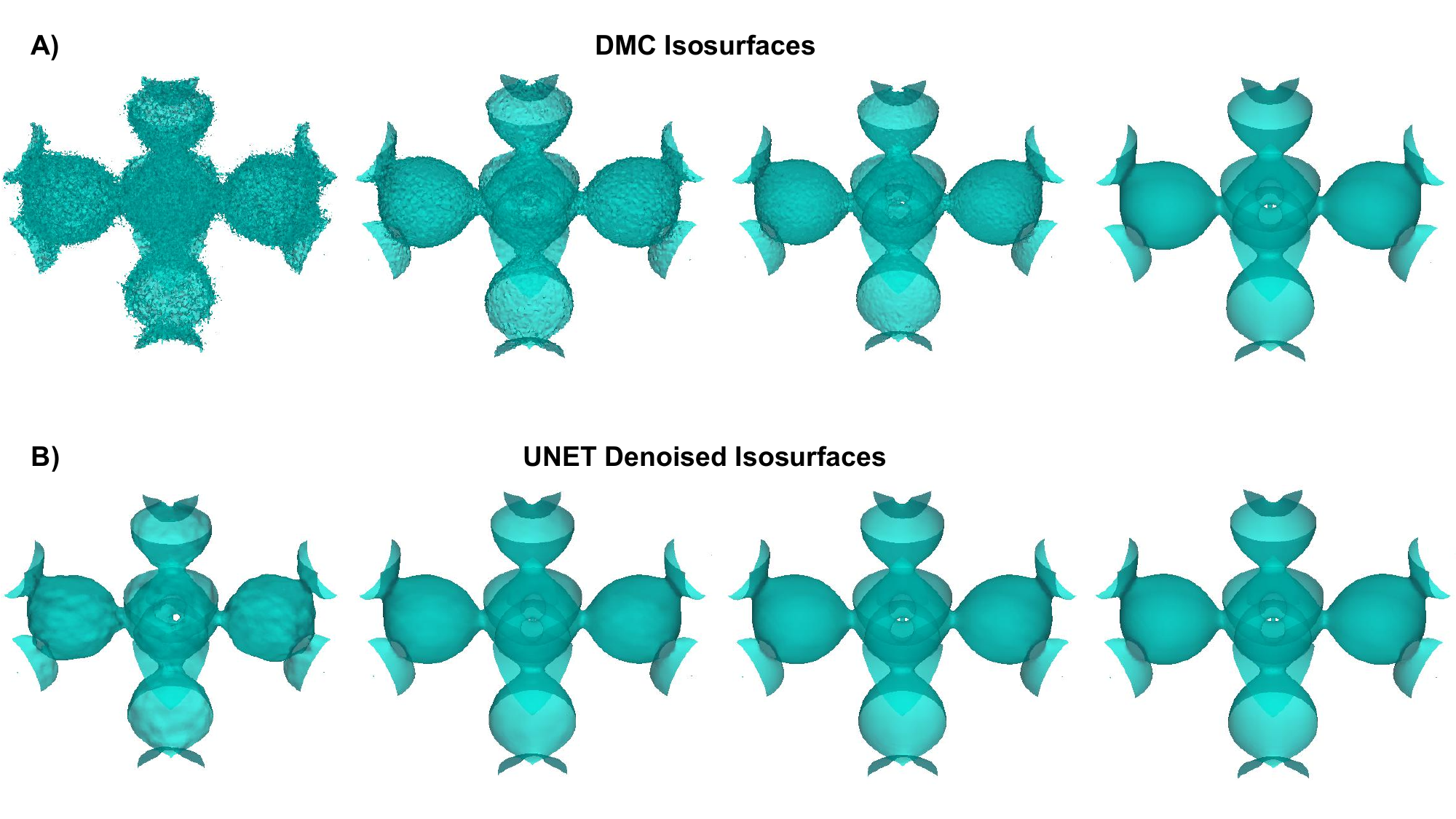}
    \caption{Panel A displays, from left to right: 292,606; 4,598,503; and 18,309,668 samples, followed by the mean reference. Panel B illustrates the corresponding denoised densities obtained using the 3D UNET, demonstrating significant noise reduction.}
    \label{fig:increasing_sample_iso}
\end{figure}
\clearpage

\section{Material Structures}

Below are the crystal structures for each material in XSF format. 
The cell is primitive in each case.

Diamond

\vspace{3mm}
\begin{minipage}{0.65\textwidth}
\flushleft
\begin{Verbatim}[fontsize=\small,frame=lines,baselinestretch=1]
 CRYSTAL
 PRIMVEC
     1.78500000    1.78500000    0.00000000
     0.00000000    1.78500000    1.78500000
     1.78500000    0.00000000    1.78500000
 PRIMCOORD
   2 1
     6   0.00000000    0.00000000    0.00000000
     6   0.89250000    0.89250000    0.89250000
\end{Verbatim}
\end{minipage}

\vspace{5mm}
Blue phosphorus

\vspace{3mm}
\begin{minipage}{0.65\textwidth}
\flushleft
\begin{Verbatim}[fontsize=\small,frame=lines,baselinestretch=1]
 CRYSTAL
 PRIMVEC
     1.65700000   -5.23870000    0.00000000
     3.31400000    0.00000000    0.00000000
     0.00000000    0.00000000    4.37536000
 PRIMCOORD
   4 1
    15   0.82850000   -1.55474139    0.35326657
    15   2.48550000   -3.68395861    2.54094657
    15   2.48550000   -1.55474139    1.83441343
    15   4.14250000   -3.68395861    4.02209343
\end{Verbatim}
\end{minipage}

\vspace{5mm}
VO$_2$

\vspace{3mm}
\begin{minipage}{0.65\textwidth}
\flushleft
\begin{Verbatim}[fontsize=\small,frame=lines,baselinestretch=1]
 CRYSTAL
 PRIMVEC
     4.55460000    0.00000000    0.00000000
     0.00000000    4.55460000    0.00000000
     0.00000000    0.00000000    2.85280000
 PRIMCOORD
   6 1
     8   1.36683546    1.36683546    0.00000000
     8   3.18776454    3.18776454    0.00000000
     8   3.64413546    0.91046454    1.42640000
     8   0.91046454    3.64413546    1.42640000
    23   0.00000000    0.00000000    0.00000000
    23   2.27730000    2.27730000    1.42640000
\end{Verbatim}
\end{minipage}

\bibliography{Bib}